\documentclass[12pt]{article}
\usepackage{amsmath}
\usepackage{txfonts,amstext,amssymb,amsfonts,color,lscape,youngtab,tikz}
\usepackage[english, USenglish]{babel}
\usepackage{graphicx}

\advance\voffset by -1.5cm
\advance\hoffset by -1.25cm
\newcommand{\tikzcircle}[2][red,fill=red]{\tikz[baseline=-0.5ex]\draw[#1,radius=#2] (0,0) circle ;}

\definecolor{darkgreen}{rgb}{0,0.6,0}
\def\thefootnote{\fnsymbol{footnote}}

\def\Tr{\,{\rm Tr}\, }

\def\tr{\,{\rm tr}\, }

\def\be{\begin{equation}}
\def\ee{\end{equation}}
\def\ba{\begin{eqnarray}}
\def\ea{\end{eqnarray}}

\newcommand{\D}{{\cal D}}

\newcommand{\N}{{\cal N}}

\newcommand{\ZZ}{\mathbb{Z}}

\newcommand{\SU}{{\rm SU}}
\newcommand{\SO}{{\rm SO}}
\newcommand{\U}{{\rm U}}

\newcommand{\nn}{{\nonumber}}

\renewcommand{\SU}{{\rm SU}}

\newcommand{\bpsiup}{\overline{\psiup}}
\newcommand{\bpartial}{\overline{\partial}}
\newcommand{\barz}{{\overline{z}}}

\def\a{a}
\def\NS{{\rm NS}}

\def\ch{{\rm ch}}
\def\br{{\mathcal X}}
\def\cp{\text{cp}}

\def\S{{\cal S}}

\newcommand{\repDid}[1]{(id, #1)}

\newcommand{\repDv}[1]{(v, #1)}

\usepackage[small,nohug,heads=vee]{diagrams}
\diagramstyle[labelstyle=\scriptstyle]
\def\N{\text{N}}
\def\D{\text{D}}

\def\q{{\mathfrak q}}
\def\reg{\rm reg}

\newcommand{\necklacetwo}[6]{
\put(3,45){\tikzcircle[black, fill=white]{0.87}}
{\put(12.5, 66.6506){\tikzcircle[black, fill=#1]{0.1}} \put(0.,
45.){\tikzcircle[black, fill=#2]{0.1}} \put(12.5,
23.3494){\tikzcircle[black, fill=#3]{0.1}} \put(37.5,
23.3494){\tikzcircle[black, fill=#4]{0.1}} \put(50.,
45.){\tikzcircle[black, fill=#5]{0.1}} \put(37.5,
66.6506){\tikzcircle[black, fill=#6]{0.1}} } }
\newcommand{\necklacethree}[9]{
\put(3,45){\tikzcircle[black, fill=white]{0.87}}
{\put(16.4495, 68.4923){\tikzcircle[black, fill=#1]{0.1}}
\put(3.34936, 57.5){\tikzcircle[black, fill=#2]{0.1}} \put(0.379806,
40.6588){\tikzcircle[black, fill=#3]{0.1}} \put(8.93031,
25.8489){\tikzcircle[black, fill=#4]{0.1}} \put(25.,
20.){\tikzcircle[black, fill=#5]{0.1}} \put(41.0697,
25.8489){\tikzcircle[black, fill=#6]{0.1}} \put(49.6202,
40.6588){\tikzcircle[black, fill=#7]{0.1}} \put(46.6506,
57.5){\tikzcircle[black, fill=#8]{0.1}} \put(33.5505,
68.4923){\tikzcircle[black, fill=#9]{0.1}}} }
\newcommand{\necklacefoura}[9]{
{\put(3,45){\tikzcircle[black, fill=white]{0.87}}} {\put(18.5295,
69.1481){\tikzcircle[black, fill=#1]{0.1}} \put(7.32233,
62.6777){\tikzcircle[black, fill=#2]{0.1}} \put(0.851854,
51.4705){\tikzcircle[black, fill=#3]{0.1}} \put(0.851854,
38.5295){\tikzcircle[black, fill=#4]{0.1}} \put(7.32233,
27.3223){\tikzcircle[black, fill=#5]{0.1}} \put(18.5295,
20.8519){\tikzcircle[black, fill=#6]{0.1}} \put(31.4705,
20.8519){\tikzcircle[black, fill=#7]{0.1}} \put(42.6777,
27.3223){\tikzcircle[black, fill=#8]{0.1}} \put(49.1481,
38.5295){\tikzcircle[black, fill=#9]{0.1}}} }
\newcommand{\necklacefourb}[3]{
{\put(49.1481, 51.4705){\tikzcircle[black, fill=#1]{0.1}}
\put(42.6777, 62.6777){\tikzcircle[black, fill=#2]{0.1}} \put(31.4705,
69.1481){\tikzcircle[black, fill=#3]{0.1}}} }
\newcommand{\necklacefivea}[9]{
\put(3,45){\tikzcircle[black, fill=white]{0.87}}
{\put(19.8022, 69.4537){\tikzcircle[black, fill=#1]{0.1}}
\put(10.3054, 65.2254){\tikzcircle[black, fill=#2]{0.1}} \put(3.34936,
57.5){\tikzcircle[black, fill=#3]{0.1}} \put(0.136953,
47.6132){\tikzcircle[black, fill=#4]{0.1}} \put(1.22359,
37.2746){\tikzcircle[black, fill=#5]{0.1}} \put(6.42138,
28.2717){\tikzcircle[black, fill=#6]{0.1}} \put(14.8316,
22.1614){\tikzcircle[black, fill=#7]{0.1}} \put(25.,
20.){\tikzcircle[black, fill=#8]{0.1}} \put(35.1684,
22.1614){\tikzcircle[black, fill=#9]{0.1}}}}
\newcommand{\necklacefiveb}[6]{
{\put(43.5786, 28.2717){\tikzcircle[black, fill=#1]{0.1}}
\put(48.7764, 37.2746){\tikzcircle[black, fill=#2]{0.1}} \put(49.863,
47.6132){\tikzcircle[black, fill=#3]{0.1}} \put(46.6506,
57.5){\tikzcircle[black, fill=#4]{0.1}} \put(39.6946,
65.2254){\tikzcircle[black, fill=#5]{0.1}} \put(30.1978,
69.4537){\tikzcircle[black, fill=#6]{0.1}}}}
\newcommand{\necklaceexamplea}[9]{
\put(3,45){\tikzcircle[black, fill=white]{0.87}}
{\put(19.8022, 69.4537){\tikzcircle[black, fill=#1]{0.1}}
\put(10.3054, 65.2254){\tikzcircle[black, fill=#2]{0.1}} \put(3.34936,
57.5){\tikzcircle[black, fill=#3]{0.1}} \put(0.136953,
47.6132){\tikzcircle[black, fill=#4]{0.1}} \put(1.22359,
37.2746){\tikzcircle[black, fill=#5]{0.1}} \put(6.42138,
28.2717){\tikzcircle[black, fill=#6]{0.1}} \put(14.8316,
22.1614){\tikzcircle[black, fill=#7]{0.1}} \put(25.,
20.){\tikzcircle[black, fill=#8]{0.1}} \put(35.1684,
22.1614){\tikzcircle[black, fill=#9]{0.1}}}}
\newcommand{\necklaceexampleb}[6]{
{\put(43.5786, 28.2717){\tikzcircle[black, fill=#1]{0.1}}
\put(48.7764, 36.1746){\tikzcircle[white, fill=#2]{0.27}} \put(49.863,
47.6132){\tikzcircle[black, fill=#3]{0.1}} \put(46.6506,
57.5){\tikzcircle[black, fill=#4]{0.1}} \put(39.6946,
65.2254){\tikzcircle[black, fill=#5]{0.1}} \put(30.1978,
69.4537){\tikzcircle[black, fill=#6]{0.1}}
\put(51.5, 40.2746){\tikzcircle[black, fill=white]{0.02}}
\put(50.7, 37.2746){\tikzcircle[black, fill=white]{0.02}}
\put(49.5, 34.2746){\tikzcircle[black, fill=white]{0.02}}}}

\begin{document}
\Yautoscale0
\addtolength{\baselineskip}{3pt}


\thispagestyle{empty}
\renewcommand{\thefootnote}{\fnsymbol{footnote}}

{\hfill \parbox{2.45cm}{
 DESY 14-028 \\
}}

\bigskip

\begin{center} \noindent \Large \bf
Chiral Primaries in Strange Metals
\end{center}

\bigskip\bigskip\bigskip

\centerline{ \normalsize \bf
 Mikhail Isachenkov$^a$
  \footnote[2]{\noindent \tt email: mikhail.isachenkov@desy.de},
 Ingo Kirsch$^a$
  \footnote[1]{\noindent \tt email: ingo.kirsch@desy.de}
and Volker Schomerus$^a$
  \footnote[1]{\noindent \tt email: volker.schomerus@desy.de}
}

\bigskip

\centerline{\it ${}^a$ DESY Hamburg, Theory Group,}
\centerline{\it Notkestrasse 85, D-22607 Hamburg, Germany}

\bigskip\bigskip

\bigskip\bigskip

\renewcommand{\thefootnote}{\arabic{footnote}}

\centerline{\bf \small Abstract}
\medskip

{\small \noindent It was suggested recently that the study of
1-dimensional QCD with fermions in the adjoint representation
could lead to an interesting toy model for strange metals and
their holographic formulation. In the high density regime, the
infrared physics of this theory is described by a constrained
free fermion theory with an emergent \mbox{${\cal N}=(2,2)$} superconformal
symmetry. In order to narrow the choice of potential holographic
duals, we initiate a systematic search for chiral primaries in
this model. We argue that the bosonic part of the superconformal
algebra can be extended to a coset chiral algebra of the form
${\cal W}_N =\SO(2N^2-2)_1/\SU(N)_{2N}$. In terms of this algebra
the spectrum of the low energy theory decomposes into a finite
number of sectors which are parametrized by special necklaces.
We compute the corresponding characters and partition functions
and determine the set of chiral primaries for $N \leq 5$.}

\newpage
\setcounter{tocdepth}{2}
\tableofcontents

\setcounter{equation}{0}
\section{Introduction}

Low dimensional examples of dualities between conformal field theories
and gravitational models in Anti-deSitter (AdS) space provide an area
of active research. There are several reasons why such developments
are interesting. On the one hand, many low dimensional critical
theories can actually be realized in condensed matter systems. As they
are often strongly coupled, the AdS/CFT correspondence might provide
intriguing new analytic tools to compute relevant physical
observables. On the other hand, low dimensional incarnations of the
AdS/CFT correspondence might also offer new views on the very working
of dualities between conformal field theories and gravitational models
in AdS backgrounds.  This applies in particular to the AdS$_3$/CFT$_2$
correspondence since there exist many techniques to solve
2-dimensional models directly, without the use of a dual gravitational
theory. Recent examples in this direction include the correspondence
between certain 2-dimensional coset conformal field theories and higher
spin gauge theories \cite{Gaberdiel:2010pz,Gaberdiel:2011zw}, see also
\cite{Creutzig,Creutzig2, Candu,Gaberdiel} for examples involving supersymmetric conformal
field theories and \cite{Gaberdiel:2014yla} for a more extensive list of the vast
literature on the subject. It would clearly be of significant interest
to construct new examples of the AdS$_3$/CFT$_2$ correspondence which
involve full string theories in AdS$_3$.

In 2012, Gopakumar, Hashimoto, Klebanov, Sachdev and Schoutens
\cite{Gopakumar} studied a two-dimensional adjoint QCD in which
massive Dirac fermions $\Psi$ are coupled to an SU$(N)$ gauge field.
The fermions were assumed to transform in the adjoint rather than the
fundamental representation of the gauge group. In the strongly coupled
high density region of the phase space, the corresponding infrared
fixed point is known to develop an \mbox{${\cal N}=(2,2)$} superconformal
symmetry. For gauge groups SU$(2)$ and SU$(3)$ the fixed points
possess Virasoro central charge $c_2 = 1$ and $c_3 = 8/3$,
respectively. These central charges are smaller that the critical
value of $c=3$ below which one can only have a discrete set of \mbox{${\cal N}=(2,2)$}
superconformal minimal models. Such theories are very well
studied. Once we go beyond $N=3$, however, the central charge $c_N =
(N^2-1)/3$ of the infrared fixed point exceeds the critical value and
the models are very poorly understood at present.  Note that the
central charge $c_N$ of these models grows quadratically with the rank
$N-1$ of the gauge group. While this is very suggestive of a string
theory dual, there exist very little further clues on the appropriate
choice of the 7-dimensional compactification manifold $M^7$ of the
relevant AdS background.

The most interesting structure inside any \mbox{${\cal N}=(2,2)$}
superconformal field theory is its chiral ring. Recall that the
\mbox{${\cal N}=(2,2)$} superconformal algebra
contains a U(1) R-charge $Q$. The latter provides a lower bound on the
conformal weights $h$ in the theory, i.e.\ physical states $\phi$ in a
unitary superconformal field theory obey the condition $h(\phi) \geq
Q(\phi)$. States in the Neveu-Schwarz sector that saturate this bound,
i.e.\ for which $h (\phi) = Q(\phi)$, are called {\em chiral primaries}.
Since chiral primaries are protected by supersymmetry, they are expected
to play a key role in discriminating between potential gravitational
duals for the infrared fixed point of adjoint QCD. More concretely, the
space of chiral primaries in the limit of large $N$ should carry
essential information on the compactification manifold $M^7$ of the
dual AdS$_3$ background.

The goal of our work is to initiate a systematic study of the chiral
ring for the models proposed by Gopakumar et al. In \cite{Gopakumar}
the partition function of the infrared fixed point was studied for
$N=2, 3$. In these two cases the chiral ring is well understood
through the relation with \mbox{${\cal N}=(2,2)$} minimal models, as we
mentioned above. The chiral primaries that are found in these two simple
models are special representatives of a larger class of {\em regular}
chiral primaries that can be constructed for all $N$. But once we pass
the critical value of the central charge, i.e.\ for $c_N > 3$, additional
chiral primaries start showing up. We shall find one example at $N=4$ and
three non-regular, or {\em exceptional}, chiral primaries for $N=5$. In
order to do so, we develop some technology that can be applied also to
larger values of $N$ and we hope that it will provide essential new
tools in order to address the large $N$ limit.

Let us briefly discuss the plan of this paper.
In the next section, we shall describe the low energy theory, identify
its chiral algebra, construct the relevant modular invariant partition
function and finally discuss the emergent \mbox{${\cal N}=(2,2)$} superconformal
symmetry. Our discussion differs a bit from the one in \cite{Gopakumar}
in that we work with a larger chiral algebra. Our algebra has the
advantage that it contains the R-current of the model. This gives us
more control over chiral primaries in the subsequent analysis. Section
3 is devoted to the representation theory of the chiral symmetry. There
we shall explain how to label its representations and how to construct
the corresponding characters. In doing so, we shall keep track of the
R-charges. The section concludes with explicit lists of representations
up to $N=5$. In section 4 we turn to the main theme of this work, the
set of chiral primaries. After explaining some general bounds on their
conformal weights we describe the set of regular chiral primaries and
study some of their properties. Finally, we construct all additional
exceptional chiral primaries for $N=4$ and $N=5$. These were not known
previously. Whether any of these additional chiral primaries survive
in the large $N$ limit remains an interesting issue for future research.

\section{The model and its symmetries}

The main purpose of this section is to review the setup described
in \cite{Gopakumar}. Starting from 2-dimensional adjoint QCD we
describe how the low energy description emerges in the  limit
of large density and strong coupling. Special attention is
paid to the chiral symmetries of the theory which are identified
at the end of the first subsection. The algebra we construct there
is a bit larger than the one that was considered in \cite{Gopakumar}.
In the second subsection we then describe how the state space of
the low energy theory decomposes into representations of left- and
right-moving chiral algebra. The section concludes with some
comments on an emergent \mbox{${\cal N}=(2,2)$} superconformal symmetry and the
role of chiral primaries for future studies of AdS duals.

\subsection{Review of the model}

The model we start with is a 2-dimensional version of QCD with fermions
in the adjoint representations, i.e.\
\begin{equation}
 {\cal L}(\Psi,A) = \Tr \left[ \overline{\Psi}(i\gamma^\mu D_\mu -m -\mu
 \gamma^0)\Psi\right]
 - \frac{1}{2g_{\text{YM}}^2 } \Tr F_{\mu\nu} F^{\mu\nu} \ .
\end{equation}
Here, $A$ denotes an SU$(N)$ gauge field with field strength $F$ and gauge
coupling $g_\text{YM}$. The complex Dirac fermions $\Psi$ transform in the
adjoint of the gauge group and $D_\mu$ denote the associated covariant
derivatives. The two real parameters $m$ and $\mu$ describe the mass and
chemical potential of the fermions, respectively.

We are interested in the strongly coupled high density regime of the
theory, i.e.\ in the regime of very large chemical potential $\mu \gg
m$ and $g_\text{YM}$. As is well known, we can approximate the
excitations near the zero-dimensional Fermi surface by two sets of
relativistic fermions, one from each component of the Fermi surface.
These are described by the left- and right-moving components of
massless Dirac fermions. At strong gauge theory coupling, the
resulting (Euclidean) Lagrangian reads
\begin{equation}
{\cal L}_{\text{eff}}(\psiup,\bpsiup,A) = \Tr
\left( \bpsiup^\ast \partial\, \bpsiup + \psiup^\ast \bpartial\, \psiup
+ A_z [\,\psiup^\ast,\psiup\,] + A_\barz [\, \bpsiup^\ast,\bpsiup\,]\right)\ .
\end{equation}
Here we have dropped the term involving the field strength $F$, using
that $g_\text{YM}\rightarrow \infty$. Upon integrating out the two
components $A_z$ and $A_\barz$ of the gauge field we obtain the
constraints
\begin{equation} \label{eq:constraint}
J(z) \ := \ [\,\psiup^\ast,\psiup\,] \ \sim\ 0  \quad , \quad
\bar J(\bar z) \ := \ [\, \bpsiup^\ast,\bpsiup\,] \ \sim \ 0 \ .
\end{equation}
These constraints are to be implemented on the state space of the
$N^2-1$ components of the complex fermion $\psiup$ such that all the modes
$J_n, n > 0,$ of $J(z) = \sum J_n z^{-1-n}$ vanish on physical states, as
is familiar from the standard Goddard-Kent-Olive coset construction~\cite{GKO}.

In order to describe the chiral symmetry algebra of the resulting
conformal field theory we shall start with the unconstrained model,
which we refer to as the {\em numerator} theory. It is based on
$M= N^2-1$ complex fermions $\psiup_\nu, \nu = 1, \dots, M$. These
give rise to a Virasoro algebra with central charge $c_\N = N^2-1$,
where the subscript $\N$ stands for numerator. We can decompose
each complex fermion into two real components $\psi^n_\nu, n=1,2,$
such that $\psiup_\nu = \psi^1_\nu + i \psi^2_\nu$. From time to
time we shall combine $\nu$ and $n$ into a single index $\alpha=
(\nu,n)$. Let us recall that the $2M$ real fermions $\psi_\alpha$
can be used to build SO($2M$) currents $K_{\alpha\beta}$ at level
$k=1$. The central charge of the associated Virasoro field coincides
with the central charge $c_\N$ of the original fermions. The
SO($2M$)$_1$ current algebra generated by the modes of $K_{\alpha\beta}$
forms the numerator in the coset construction.

In order to describe the {\em denominator}, i.e.\ the algebra generated
by the constraints \eqref{eq:constraint}, we need to recall a second way
in which our fermions $\psi^n_\nu$ give rise to currents. According to
the usual constructions, we can employ the representation matrices of
the adjoint representation to build two sets of SU($N$)
currents at level $k=N$. These currents will be denoted by $j^n_\nu$
with $\nu=1,\dots,M$ and $n=1,2$. The currents $J$ that were introduced
in eq.\ \eqref{eq:constraint} are obtained as $J_\nu = j^1_\nu + j^2_\nu$.
The chiral SU($N$) currents $J_\nu$ form an affine algebra at level
$k=2N$. Through the Sugawara construction we obtain a Virasoro algebra
with central charge $c_\D = 2(N^2-1)/3$, where the subscript $\D$ stands
for denominator. Now we have assembled all the elements that are
needed in defining the coset chiral algebra
\begin{equation}\label{eq:W2N}
{\cal W}_N := \text{SO}(2N^2-2)_1 / \text{SU}(N)_{2N}\ .
\end{equation}
The parameter $N$ keeps track of the gauge group SU($N$). The algebra
${\cal W}_N$  is a key element in our subsequent analysis. It is larger
than the chiral symmetry considered in \cite{Gopakumar} which uses the
subalgebra SU$(N)_N \times$ SU$(N)_N \subset$ SO$(2N^2-2)_1$ to encode
symmetries of the numerator theory.

\subsection{Modular invariant partition function}

The coset algebra ${\cal W}_N$ describes the chiral symmetries
of our model. Consequently, the partition function must decompose into
a sum of products of characters for the left- and right chiral symmetry.
These characters will be discussed in much detail below. The aim of this
section is to explain how they are put together in order to construct
the partition function of the coset model.

We shall begin with a few simple comments on the numerator theory. As
we reviewed above, its state space carries the action of a chiral SO($2M$)
algebra at level $k=1$. This current algebra possesses four sectors which
are denoted by {\it id,v,sp} and {\it c}, respectively. When decomposed into
the associated characters, the partition function takes the form
\begin{align}
Z^{\N}(q,\bar q) &=\  |\chi_{\text{\it id}} ^{\N}|^2+ |\chi_v ^{\N}|^2
+  |\chi_{\text{\it sp}} ^{\N}|^2 +  |\chi_{\text{\it c}} ^{\N}|^2 \ =
\ M^\N_{AB}\,  \chi_{A}^{\N}(q) \bar
\chi_{B}^{\N}(\bar q) \, .\label{Znum}
\end{align}
The labels $A,B$ on the right hand side run through $A,B$ = {\it
id,v,sp} and {\it c} and $M^\N_{AB}$ are integers which are defined
through the expression on the left hand side. Explicitly, these
integers are given by $M^\N_{AB}={\rm diag}(1,1,1,1)$.

Now we need to describe a similar set of integers $M^\D_{ab}$ for the
denominator theory. This is obtained from the D-type modular invariant
partition function for the SU($N$)$_{2N}$ Wess-Zumino-Witten model.
Before we can spell it out, we need a bit of notation. To begin with,
we introduce the set ${\cal J}_N$ of SU($N$) weights $a = [\lambda_1,\dots,
\lambda_{N-1}]$ subject to the condition $\sum_{s=1}^{N-1} \lambda_s \leq 2N$.
These label sectors of the SU($N$) current algebra at level $k=2N$.

On ${\cal J}_N$ we can define an action of $\mathbb{Z}_{N}$ such that
\begin{align} \label{eq:SC}
\gamma([\lambda_1,\dots,\lambda_{N-1}]) = [2N-\sum_{s=1}^{N-1}\lambda_s,\lambda_1,\dots,
\lambda_{N-2}]\
\end{align}
for the generator $\gamma \in \mathbb{Z}_{N}$. Obviously, $\gamma$
maps elements $a \in {\cal J}_N$ back into ${\cal J}_N$ and it obeys
$\gamma^{N} =${\it id}.

In addition, we can also construct a map $h^\D:{\cal J}_N \rightarrow
\mathbb{R}$ that assigns a conformal weight $h^\D(a)$ to each sector
$a \in {\cal J}_N$. The weight is given by\footnote{The quadratic Casimir
 of an $\SU(N)$ representation $a$  is given by
\begin{align}
C_2(a)
&=\frac{1}{2} \left[-\frac{n^2}{N} + n N  +
\sum_{i=1}^r (  l^2_i + l_i - 2i l_i) \right] \,, \nonumber
\end{align}
where $n=\sum_{i=1}^r l_i$ is the total number of boxes in the
corresponding Young tableau, and $l_i = \sum_{s = i}^{N-1} \lambda_s$
($i=1,..., r$) denotes the length of the $i$th row.}
\begin{equation}
h^\D(a) =   \frac{C_2(a)}{3N}\,.
\end{equation}

With the help of the map $\gamma: {\cal J}_N \rightarrow {\cal J}_N$
and the weight $h^\D: {\cal J}_N \rightarrow \mathbb{R}$ we can
finally define the so-called monodromy charge
\begin{align}\label{monocharge}
Q_\gamma(a) \equiv h^\D(\gamma(a))-h^\D(a) \ \mbox{mod} \ 1 .
\end{align}
Now we have collected all the ingredients we need in order to spell
out the desired D-type modular invariant partition function of the
SU$(N)_{2N}$ Wess-Zumino-Witten model,
\begin{align}
Z^\D(q,\bar q) \ = \ \sum_{\{a\}; Q_\gamma(a) \equiv 0}
\frac{N}{N_a} \vert \sum_{b \in \{a\}} \chi_b \vert^2
\ = \ \sum_{ab} M^\D_{\a b}\,  \chi_{\a}^{\SU(N)_{2N}}(q) \bar
\chi_{b}^{\SU(N)_{2N}}(\bar q)
\ .  \label{Zdenom}
\end{align}
The first summation is over orbits $\{a\}$ of weights for the affine
$\SU(N)_{2N}$ under the action \eqref{eq:SC} of the identification
current $\gamma$. The length of a generic orbit agrees with the size
$N$ of the gauge group SU$(N)$. Some orbits $\{a\}$, however, possess
fixed points so that their length $N_a = N_{\{a\}}$ can be a nontrivial
divisor of $N$. For more details on simple current modular invariants
see \cite{Yankielowicz}.

From eq.\ \eqref{Zdenom} we can read off the integer coefficients
$M^\D$ of the decomposition \cite{Bernard},
\begin{align}
M^\D_{a b} & = \ \left\{
\begin{array}{ll}
 \ \sum_{p=1}^{N} \delta_{\a, \sigma^p(b)} &\textmd{if }\,
 t(a)=0 \mod N\\[4mm]
 \ 0 &\textmd{otherwise}
\end{array}
 \right.  ,
\end{align}
where $t(a)=\sum_{s=1}^{N-1} s\,\lambda_s $ is the $N$-ality of an
SU$(N)_{2N}$ representation $a = [\lambda_1, ..., \lambda_{N-1}]$ and
$\sigma^r(a) = 2N w_r + c^r(a), r=1,...,N$, are associated with group
automorphisms of SU$(N)_{2N}$. They are defined in terms of the
fundamental weights $w_r$ and the Coxeter rotations $c^r(w_i) =
w_{i+r}- w_r$, see \cite{Bernard, Aldazabal} for more details and \cite{Cappelli, Gannon}
for $M^\D$ at $N=2,3$. Note
that the $N$-ality constraint coincides with the condition of vanishing
monodromy charge that was built into eq.\ \eqref{Zdenom}.

We are now prepared to construct a modular invariant that is associated
with our coset model. In fact, following the standard procedures in coset
conformal field theory we are led to consider
\begin{align}\label{partfualt}
\tilde Z_N(q,\bar q) = \frac{1}{N^2} \sum_{ABab} M^\D_{\a b} M^\N_{AB}
\br^{\cal W}_{(A,\a)}(q)
\bar{\br}^{\cal W}_{(B,b)}(\bar q)\, .
\end{align}
The summation runs over the same range as in eqs.\ \eqref{Znum} and
(\ref{Zdenom}). The functions $\br$ are so-called branching functions.
We will define and construct them in the next section. For most values
of the label $(A,\a)$, the branching function $\br$ is a character $\chi$
of an irreducible representation of ${\cal W}_N$. More precisely, one finds
that
\begin{equation}
\br^{\cal W}_{(A,\a)}(q) = \chi^{\cal W}_{(A,\a)}(q) \quad \text{ when }
\quad N_\a = N \ ,
\end{equation}
i.e.\ when the orbit $\{a\}$ of $\a$ under the action of $\gamma \in
\mathbb{Z}_N$ consists of $N$ elements. The orbit $\{0\}$ of the vacuum
representation is always such a long one. Consequently, in order for
the vacuum to contribute with unit multiplicity, we had to divide the
sum in eq.\ \eqref{partfualt} by $N^2$.  But this is a dangerous
division. In order to see the problem, let us insert $M^\N_{AB}$ and
$M^\D_{\a b}$ as in eqs.\ (\ref{Znum}) and (\ref{Zdenom}),
respectively. Then our modular invariant \eqref{partfualt} reads
\begin{align}
\tilde Z_N(q,\bar q) = \sum_A \sum_{\{a\}, Q_\gamma(\a)=0} \frac{N_\a}{N}
\vert \br^{\cal W}_{(A,\a)} (q)\vert^2
\,. \label{partfu}
\end{align}
Here, we sum over orbits $\{a\}$ instead of SU$(N)$ representations $a$
with vanishing monodromy charge. For short orbits we have $N_a <
N$ so that the corresponding branching functions are divided by a
non-trivial integer. Typically, one finds that these fractions are not
compensated by corresponding multiplicities in the branching functions
so that the modular invariant \eqref{partfualt} possesses non-integer
coefficients. This problem is of course well known and may be overcome
by a process known as fixed point resolution, see
\cite{Yankielowicz,Fuchs}. In the case of short orbits, i.e.\ when
$N_a \neq N$, the branching function $\br^{\cal W}_{(A,\a)}$ turns out
to decompose into a sum of ${\cal W}_N$ characters $\chi^{\cal
W}_{(A,\a,m)}$ for irreducibles labeled by $m$. General formulas
for such decompositions exist only for some coset chiral algebras, see
e.g.\ \cite{Fuchs}.  Experience shows that the characters $\chi^{\cal
W}_{(A,\a,m)}$ can be used as building blocks for modular invariants
$Z^\text{res}_N$ such that
\begin{equation} \label{ZtZrZ}
 Z_N(q) = \tilde Z_N(q) + Z^\text{res}_N(q) \
\end{equation}
has integer coefficients only. $Z_N$ can therefore be interpreted as the
partition function of the system. To spell out details, we shall mostly
assume that $N$ is a prime number. Under this condition, the sectors
$(A,[2,2,\dots,2])$ turn out to generate the only short orbits and
the resolution process can be spelled out explicitly. Following a
recipe first described in \cite{Yankielowicz} we define
\begin{align}
\chi^{\cal W}_{(A,a_\ast,m)}(q) & = \frac{1}{N}
\left(\br^{\cal W}_{(A,a_\ast)}(q) + d_{(A,m)} \right) \,,\label{replacement1}
\\[2mm]
\mbox{where}\ \  \ d_{(A,m)} & \in \mathbb{Z} \quad  \mbox{ with }
\ \ \sum_{m=1}^N d_{(A,m)} = 0 \ , \label{replacement}
\end{align}
$m = 1, \dots, N$ and $A$ runs through its four possible
values, as usual. We also introduced the shorthand $a_\ast =
[2,2,\dots,2]$. Note that the proposed characters indeed sum
up to the branching functions. 
For $N>2$ we propose the
following values for $d_{(A,m)}$:\footnote{For $N=2$, we have $d_{(sp,1)}=-1$, $d_{(sp,2)}=1$
and $d_{(c,1)}=1$, $d_{(c,2)}=-1$, all others are zero.}
\begin{equation}
\begin{aligned}
d_{(id,1)}&= N-1\,,  \hspace{3.5cm}  &d_{(id,p)}&= -1\,,\\
d_{(v,1)}&= 0\,,  &d_{(v,p)}&= 0\,,\\
d_{(sp,1)}&= -\frac{(N-1)^2}{2}\,,  &d_{(sp,p)}&= \frac{N-1}{2}\,,\\
d_{(c,1)}&= \frac{N^2-1}{2}\,,  &d_{(c,p)}&= -  \frac{N+1}{2}\,,
\end{aligned}
\end{equation}
$p=2,...,N$. Given these characters we can now
construct $Z^\text{res}$ through
\begin{equation}\label{Zres}
 Z^\text{res}_N(q)\  = \ \frac{1}{N^2} \sum_A \sum_{m=1}^N
 d_{(A,m)}^2  = \frac{N^2+3}{2}  \frac{N-1}{N} 
\end{equation}
for $N$ prime.  Since $Z^\text{res}$ is a constant, it is
obviously modular invariant. In addition, if we add this
term to the modular invariant $\tilde Z_N$ we obtain an
expression in which squares of characters are summed with
integer coefficients,
\begin{equation} \label{partitionfct}
Z_N(q) = \sum_A \sum_{
\substack{\{a\}, a \neq a_\ast \\  Q(a)=0}
}
\vert \chi^{\cal W}_{(A,\a)} (q)\vert^2 +
\sum_A \sum_{m=1}^N \vert \chi^{\cal W}_{(A,a_\ast,m)}
(q)\vert^2\ .
\end{equation}
Here, the first summation is over all orbits of length
$N$ with vanishing monodromy charge. Of course our
assumption that $N$ be prime is crucial for the validity
of the expression \eqref{partitionfct} for the partition
function of our model.

\subsection{Comments on superconformal symmetry}

According to the usual Goddard-Kent-Olive (GKO) construction \cite{GKO}, the
chiral algebra ${\cal W}_N$ contains a Virasoro field whose central
charge is given by the difference of the central charges in the
numerator and the denominator,
\begin{align}  \label{eq:c}
c = c({\cal W}_N) = c_\N - c_\D = N^2-1 - \frac{2}{3}(N^2-1) =
\frac{1}{3}(N^2-1)\ .
\end{align}
Of course, the coset chiral algebra contains many more fields. To be
precise, any element of the numerator algebra that has trivial
operator product with respect to the denominator currents makes it
into our algebra ${\cal W}_N$. In the case at hand, the condition is
also satisfied by the $\U(1)$ current
\begin{align} \label{eq:current}
J(z) =\textstyle \frac{1}{3} \sum_{\nu,\mu} \psi^1_\nu(z) \psi^2_\mu(z) \kappa^{\nu\mu} \
\end{align}
where $\kappa^{\nu\mu}$ denotes the Killing form of $\SU(N)$.
This $\U(1)$ current will play a very important role.

It was observed in \cite{Boucher} that the conformal symmetry is
actually enhanced to a \mbox{${\cal N}=(2,2)$} superconformal one. This means that the
state space admits the action of fermionic generators $G^\pm$ and an
additional U(1) current $J$.  While $G^\pm$ are not contained in our
chiral algebra ${\cal W}_N$, the U(1) current is. In fact, it is
precisely the current we found in the previous paragraph.

Let us recall that in models with \mbox{${\cal N}=2$} supersymmetry there
is an important subset of fields, namely the (anti-)chiral primaries. By
definition, these correspond to states in the Neveu-Schwarz sector of
the theory, i.e.\ with $A =${\it id,v}, such that $h = |Q|$ where $Q$
denotes the U(1) charge and $h$ the conformal weight. Chiral primaries
have many interesting properties. In particular, they give rise to the
so-called chiral ring. In addition the space of chiral primaries is
protected under deformations preserving the \mbox{${\cal N}=(2,2)$}
superconformal symmetry.  Therefore, it can serve as a ``fingerprint''
of our model.

As we discussed in the introduction, chiral primaries should play an
important role when it comes to identifying the AdS dual of the
superconformal field theory we are dealing with. As we pointed out
in the introduction, AdS duals of 2-dimensional (super-)conformal
field theories have recently attracted quite some attention. In the
existing examples, the central charge is linear in $N$ and the dual
model is a higher spin theory in AdS$_3$. The case we are dealing with
here is different: The central charge \eqref{eq:c} is quadratic
in $N$ and hence standard arguments would suggest a richer dual
model which is described by a full string theory in AdS$_3$ rather
than a higher spin theory. The identification of this string theory
would be significant progress. Clearly, the chiral primaries could
play a central role in identifying the string dual.

\setcounter{equation}{0}
\section{Representations of the chiral symmetry}

In the previous section we identified the chiral symmetry algebra ${\cal
W}_N$ of the coset model. Our next aim is to develop the
representation theory of this chiral symmetry. In the first subsection
we shall provide several different ways to think about the pairs
$(A,\a)$ that label non-trivial branching functions of our chiral
algebra ${\cal W}_N$.  Then we explain how to obtain the branching
functions from the characters of the numerator and the denominator
theory. We have worked out the first few terms in the expansion of
these characters for all representations with $N \leq 5$. The results
are sketched in the third subsection, at least to the extent to which
they are needed later on. More details may be found in appendix B.

\subsection{Labeling of orbits}

The labels $(A, a)$ of ${\cal W}_N$ branching functions have been
described in the previous section already. Let us recall that $A$ runs
through the four values $A =${\it id,v,sp,c}. The range of $a$ was
a little more difficult to state. It should be taken from the set
${\cal J}^0_N$ of SU$(N)_{2N}$ labels $a$ with vanishing monodromy charge
 $Q_\gamma(a)= 0$, see eq.\ \eqref{monocharge}. 
Since branching functions are invariant under the action  \eqref{eq:SC}
of the identification group $\mathbb{Z}_{N}$, we only need to pick
one representative $a$ from each orbit $\{a\} \in {\cal O}_N = {\cal J}_N^0/\ZZ_N$. 
In the next two subsections we will develop an approach that allows to
enumerate the branching functions of ${\cal W}_N$ more systematically.

\subsubsection{Solving the zero monodromy condition}\label{integersec}

Our first task is to describe those labels $a$ that solve the
 $Q(a)=0$ condition, i.e.\ elements $a \in {\cal J}^0_N$.
Our claim is that elements $a$ of ${\cal J}^0_N$ are in one-to-one
correspondence with pairs of SU$(N)$ Young diagrams $Y'$ and $Y''$
with equal number $n' = |Y'|= |Y''|$ of boxes satisfying
\begin{equation} \label{condgeneration}
r' + c'' \leq N \quad ,\quad r'' + c' \leq 2N \
\end{equation}
where $r',r''$ and $c',c''$ denote the numbers of rows and columns
of $Y',Y''$, respectively. Let us denote the row lengthes of the
Young diagrams $Y'$ and $Y''$ by\footnote{The row lengths $l'_i$
($i=1,..., r'$) are related to the Dynkin labels $\lambda'_s$ by
$l'_i = \sum_{s = i}^{N-1} \lambda'_s$ (and similarly for $l_i''$).}
$$ Y' = (l'_1,\dots,l'_{r'}) \quad , \quad  Y'' = (l''_1,\dots,l''_{r''})
\ .$$
Here, we arrange the $l'_i$ and $l''_i$ in decreasing order, i.e.\
$l'_i \geq l'_{i+1}$ etc. so that the largest entries are $l_1'  =
c'$ and $l_1'' = c''$. From these two Young diagrams we can build
a new diagram $Y = (l_1, \dots, l_{N-1})$ through\footnote{This extends 
the construction of \cite{Schwimmer}.}
\begin{align}
l_i = \left\{
\begin{array}{lll}
r'' +l'_i  &\rm for& i=1, ..., r'\\[2mm]
r'' &\rm for & i=r' +1, ..., N-l''_1\\[2mm]
r'' -k &\rm for & i=N-l''_k+1, ..., N-l''_{k+1} \,, \quad k=1, ..., r'' -1 \\[2mm]
0 &\rm for & i=N-l''_{r''}+1, ..., N-1\,.
\end{array}
\right. \label{generation}
\end{align}
As will be shown in appendix~C, the total number $|Y|$ of boxes in $Y$
is $n= r'' N$ and the value of the quadratic Casimir is
\begin{align}
C_2(a) = C_2(Y) = n' N + C_2(Y') - C_2(Y'') \,, \label{C2generalized}
\end{align}
where $C_2(Y')$ and $C_2(Y'')$ are the quadratic Casimirs of the
$\SU(N)$ representations associated with $Y'$ and $Y''$, respectively.
Formula \eqref{C2generalized} follows directly by substituting eq.\
(\ref{generation}) into the definition of the quadratic Casimir
invariant of $\SU(N)$.

We will not prove the parametrization of ${\cal J}^0_N$ through pairs
$(Y',Y'')$ here. But let us make a few comments at least. To begin
with, the first constraint in eq.\ (\ref{condgeneration}) is a necessary
condition for $Y=(l_1, ..., l_{N-1})$ to be a representation of
$\SU(N)$ while the second constraint ensures that it is also a
representation of the affine group $\SU(N)_{2N}$. So, the two
constraints together ensure that $Y$ corresponds to a representation
of the SU$(N)$ current algebra at level \mbox{$k=2N$}, i.e.\ to an
element of our set ${\cal J}_N$.  It is not difficult to show that the
representations $Y$ also obey the zero monodromy condition or,
equivalently, the $N$-ality condition $\sum_i^{N-1} i \lambda_i = \sum_i
l_i = 0 \mod N$. Since \mbox{$|Y|=n=r''
N$} for any pair $(Y', Y'')$, $|Y|=\sum_i l_i$ is always a multiple of $N$ and
the $N$-ality condition trivially holds true, so \mbox{$Y \in {\cal
J}^0_N$}. Checking that the representations $Y$ give the complete set
${\cal J}^0_N$ would require some more work.
We have checked with computer algebra up to \mbox{$N=7$} that the
prescription (\ref{generation}) indeed provides the complete set of
$\SU(N)_{2N}$ representations with vanishing monodromy charge.

Once we accept this parametrization of sectors with vanishing
monodromy charge through pairs $(Y',Y'')$ of Young diagrams, it is
easy to count. Indeed, we have checked up to $N=9$ that the total
number of elements in ${\cal J}^0_N$ is given by the series A082936 in
\cite{OEIS},
\begin{align}
 |{\cal J}^0_N| =\frac{1}{3N} \sum_{n|N}  \varphi(N/n) \,
 \binom{ 3n}{ n}   \,, \label{series}
\end{align}
where $\varphi(n)$ is Euler's phi function. When $N=7$, for example,
we obtain $|{\cal J}_7^0| = 5538$ representations of the affine
algebra $\SU(7)_{14}$ with vanishing monodromy charge.

\subsubsection{Necklace representation of orbits}

The partition function $Z_N$ involves a summation over orbits $\{a\}$ of
weights $a \in {\cal J}_N^0$ for the affine algebra $\SU(N)_{2N}$ under
the action \eqref{eq:SC} of the identification current $\gamma$. The
right way to proceed is therefore to group the elements of ${\cal J}^0_N$
into orbits $\{a\}$. It turns out that the orbits possess a nice
representation in terms of {\em necklaces} with  $N$ black and $2N$
white beads. A necklace is constructed from the affine Dynkin labels
$[\lambda_0, \lambda_1, ..., \lambda_{N-1}]$ of any representation within
a given orbit. We stated the relation between the row lengthes $l_i$ and
the Dynkin labels $\lambda_i$ in the previous section.
The additional entry $\lambda_0$ of the affine Dynkin label is simply
given by $\lambda_0 = 2N-\sum_{i=1}^{N-1} \lambda_i$. Necklaces are
direct graphical representations of the affine Dynkin labels. The
entries of the affine Dynkin label determine the number of white beads
which are separated by the black beads, i.e.\ the structure of a
necklace is: $\lambda_0$ white beads, black bead, $\lambda_1$ white
beads, black bead, etc.
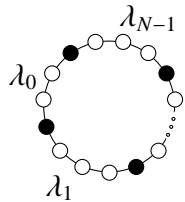
\begin{center}
\begin{picture}(80,90)
\necklaceexamplea{white}{ black}{ white}{ white}{ black}{ white}{ white}{ white}{ black}
\necklaceexampleb{white}{ white}{ white}{ black}{ white}{ white}
\put(-10,56){$\lambda_0$}
\put(32,78){$\lambda_{N-1}$}
\put(4,14){$\lambda_{1}$}
\end{picture}
\vspace{-0.5cm}
\end{center}
A necklace represents the whole orbit $\{a\}$, since the action of the
identification current $\gamma$  corresponds  to a  rotation of the
necklace but not a modification of  the necklace itself. The
identification of orbits with necklaces enables us to find a
simple formula for the number
\begin{align} \label{noorb}
|{\cal O}_N| =
\sum_{N_a|N} \frac{1}{3N_a^2} \sum_{n|N_a} \mu(N_a/n)\, { 3n \choose n }
\end{align}
of orbits ${\cal O}_N ={\cal J}_N^0/\mathbb{Z}_N$. Here, $\mu$ denotes
the classical M\"obius function.

Let us stress that the number $4|{\cal O}_N|$ counts the number of
inequivalent branching functions and not the number of representations
of our chiral algebra. If we assume that a branching function associated
to an orbit of length $N_a$ can be decomposed into characters of $N/N_a$
inequivalent representations, then the number of ${\cal W}_N$ sectors is
given by
\begin{align}
|{\cal R}^{\cal W}_N| = 4 \sum_{N_a|N} \frac{N}{N_a} \frac{1}{3N_a^2}
\sum_{n|N_a} \mu(N_a/n)\, { 3n \choose n } \nn \, .
\end{align}
This formula produces the correct results at least when $N$ is prime.
The factor of $4$ in front of the sum stems from the summation over
$A$.

\subsection{Representations and characters}\label{secreps}

Having parametrized and counted the orbits of $(A,a)$ we will discuss
the associated branching functions and the closely related characters
of the chiral algebra ${\cal W}_N$ in more detail. By definition, the
character of a ${\cal W}_N$  representation $R$ is obtained through
\begin{align}
\label{eq:cosetchi}
\chi^{\cal W}_{R}(q,x) = \tr_{R} \left( q^{L^{\cal W}_0  -
\frac{c}{24}} x^{2J_0} \right)
\end{align}
where $L_0^{\cal W}$ denotes the zero mode of the coset Virasoro
algebra and $J_0$ is the zero mode of the current \eqref{eq:current}.
The subscript $R$ refers to the choice of a representation of the
chiral algebra ${\cal W}_N$.

As usual in coset conformal field theory we can obtain branching
functions by decomposing the characters
$\chi^\N(q,x,{\color{black}z_i})$ of the numerator $\N \equiv
\SO(2M)_1$ into characters $\chi^\D(q, {\color{black}z_i})$ of the
denominator $\D \equiv \SU(N)_{2N}$,
\begin{align} \label{eq:dec}
\chi^\N_A(q,x, {\color{black}z_i}) = \tr_A  \left( q^{L^\N_0 - \frac{c}{24}}
 x^{2J_0}  {\color{black}  \prod_{i=1}^{N-1} z_i^{H_{i 0}}} \right) =
 \sum_{A,a} \br^{\cal W}_{(A,a)}
(q,x) \chi^\D_a(q, {\color{black}z_i})\ .
\end{align}
Here we have twisted the characters of the numerator free fermion model
with the zero modes $H_{i 0}$ of the Cartan currents of $\D$
in Chevalley basis (constructed from fermions). By the
very definition of the coset chiral algebra ${\cal W}_N$ this implies
that states of the coset algebra possess  vanishing $H_{i 0}$
charge. In other words, all the dependence on the variables $z_i$
on the right hand side of the previous equation is contained in the
characters $\chi^\D$ of the denominator algebra SU$(N)_{2N}$. The
summation in eq.\ (\ref{eq:dec}) runs over representations $a$ of the
denominator algebra, i.e.\ over weights $a = [\lambda_1,\dots,
\lambda_{N-1}]$ of $\SU(N)$ subject to the condition $\sum_{s=1}^{N-1}
\lambda_s \leq 2N$. Note that the generator of the latter carry no
charge with respect to the $\U(1)$ current $J(z)$ so that the
corresponding characters are independent of $x$. The label $A$
runs through the four sectors $A = \text{\it id},v,\text{\it
sp,c}$ of the $\SO(2M)$ current algebra at level $k=1$, as before.

Let us note the following fundamental properties of the branching
functions introduced in eq.\ \eqref{eq:dec},
\begin{align}
\br^{\cal W}_{(A,a)}(q,x) & = 0 \hspace{2.45cm} \mbox{ if }
 Q_\gamma(a) \neq 0  \ , \label{Qequal0}\\[2mm]
\br^{\cal W}_{(A,a)}(q,x) & = \br^{\cal W}_{(B,b)}(q,x)
\qquad \mbox{ if } \ B = A \ , \ b = \gamma^n(a)
\end{align}
for some choice of $n$. Using these two properties, we can
rewrite eq.\ \eqref{eq:dec} in the form
\begin{align} \label{brSU}
  \chi_{A}^{\N}(q,x, {\color{black}z_i})  &=
  \sum_{\{a\}, Q_\gamma(a)=0} \br^{\cal W}_{(A,a)}(q,x)
  S_{\{a\}}(q, {\color{black}z_i})
\end{align}
where the sum extends over orbits $\{a\}$ of denominator labels under
the identification current $\gamma$ whose monodromy charge vanishes
and we defined
\begin{align} \label{Sdef}
  S_{\{a\}}(q, {\color{black}z_i}) = \sum_{b \in \{a\}} \chi^{\D}_b(q,
  {\color{black}z_i})\ .
\end{align}
In order to progress, we must now insert explicit formulas for the
various characters. The functions on the left hand side of eq.\
\eqref{brSU} are actually very easy to construct from the free
fermion representation which gives
\begin{align}
&\chi_{\text{\it id}}^{\N}(q,x, {\color{black}z_i}) \pm \chi_{v}^{\N}(q,x, {\color{black}z_i}) \nonumber\\
&\qquad\qquad =\prod_{X \in \SU(N)}\left( q^{-1/24}  \prod_{n=1}^\infty (1 \pm x^{1/3} {\color{black}z^{\alpha(X)}} q^{n-1/2})
(1 \pm x^{-1/3}  {\color{black}z^{\alpha(X)}}  q^{n-1/2}) \right) \,,\label{charSO1} \\[2mm]
&\chi_{\text{\it sp}}^{\N}(q,x, {\color{black}z_i} ) \pm \chi_{\text{\it c}}^{\N}(q,x, {\color{black}z_i} ) \nonumber\\
&\qquad\qquad= \prod_{X \in \SU(N)}
 \Big( q^{1/12} x^{1/6} \prod_{n=1}^\infty (1 \pm x^{1/3}{\color{black}z^{\alpha(X)}} q^{n})(1 \pm x^{-1/3} {\color{black}z^{\alpha(X)}}q^{n-1}) \Big) \,,
\label{charSO2}
\end{align}
where
\begin{align}
&{\color{black}z^{\alpha(X)}\equiv z_1^{\alpha_1(X)} z_2^{\alpha_2(X)} \cdots z_{N-1}^{\alpha_{N-1}(X)} }\,
\end{align}
and $\alpha(X)=(\alpha_1(X), ...,\alpha_{N-1}(X))$ is a root vector of
$\SU(N)$.  Of course, we can obtain explicit formulas for the
characters $\chi^\N_A, A = ${\it id,v} by taking the sum and
difference of the expressions in the first line.

Characters of the denominator algebra SU$(N)_{2N}$ are a little bit
more complicated but of course also well known. In terms of the string
functions $c^{b}_{\lambda}(q)$ of the denominator theory, the
characters can be written as
\begin{align}
  \chi^{\D}_b(q, {\color{black}z_i}) &=
 \sum_{\lambda \in P/k M} c^{b}_{\lambda}(q) \Theta_\lambda(q, z_i) \,,\\
\Theta_\lambda(q, z_i) &= \sum_{\beta^\vee \in M^\vee}  q^{\frac{k}{2} | \beta^\vee + \lambda/k|^2}
\prod_{i=1}^{N-1} z_i^{k (\beta^\vee + \lambda/k, \alpha^\vee_i) } \,,
\end{align}
where $P$ and $M$ ($M^\vee$) denote the weight and (co)root lattice
of $\SU(N)$, respectively, and $\alpha^\vee_i$ are the simple coroots
of $\SU(N)$.  By comparing the $q$-expansion of the right-hand side of
equation (\ref{brSU}) with that of expressions
(\ref{charSO1},\ref{charSO2}), we find the $x$-dependence of the
branching functions order by order in $x$ and~$q$.

\begin{figure}
\begin{center}
\begin{tabular}{ccccc}
 ($\bullet$, $\bullet$) & \quad $\longrightarrow$ \quad  & $\bullet$ &\quad \quad  &
 $[0]$\\[6mm]
 (\yng(1), \yng(1) ) & $\longrightarrow$ & \yng(2) & & $[2]$\\[6mm]
 (\yng(2), \raisebox{-0.46cm}{\yng(1,1)} ) & $\longrightarrow$ & \yng(4) & & $[4]$
\end{tabular}
\end{center}
\caption{Pairs of Young diagrams $(Y',  Y'')$ inducing $Y \in {\cal J}_N^0$ for $N=2$.} \label{FigVNeq2}
\end{figure}

\subsection{Examples with small $N$}\label{secnecklaces}

In order to illustrate the constructions we outlined above and to
prepare for our search of chiral primaries, we want to work out some
explicit results with $N \leq 5$. Let us recall that the central
charge of the models with $N=2$ and $N=3$ satisfies $c_N < 3$ so
that these two models are part of the minimal series of \mbox{${\cal N}=(2,2)$}
superconformal theories. The other two cases, $N=4$ and $N=5$,
however, are outside this range and hence our results here are new.

\subsubsection*{$\mathbf{N=2}$}

For $N=2$, there are $|{\cal J}^0_2|=3$ representations of SU$(2)_4$
with vanishing monodromy charge. Such representations can be
constructed from pairs of Young diagrams $( Y', Y'')$ by eq.\
(\ref{generation}), as shown in figure~\ref{FigVNeq2}. Under the
action of $\gamma \in \mathbb{Z}_2$ these representations form two
orbits. The first one is long, i.e.\ $N_{\{0\}} =N =2$ and it consists
of $\{[0], [4]\}$. There is a second orbit of length $N_{\{2\}}=1$
which is given by~$\{[2]\}$.

In the case at hand, it is  actually possible to derive   explicit
expressions for the branching functions $\br^{\cal W}$ from the general
decomposition formula (\ref{brSU}), see appendix~\ref{appA},
\begin{equation}
\begin{aligned}
\br^{\cal W}_{(id\pm v,[0])}(q,x)&= \frac{1}{\eta(q)} \sum_{n\in \mathbb{Z}} (\pm 1)^n q^{\frac{3}{2}n^2} x^n \,, \\
\br^{\cal W}_{(id\pm v,[2])}(q,x)&= \frac{1}{\eta(q)} \sum_{n\in \mathbb{Z}} (\pm 1)^{n+1} q^{\frac{3}{2}\left(n+\frac{1}{3}\right)^2}
(x^{n+\frac{1}{3}}+x^{-(n+\frac{1}{3})})\,, \\
\br^{\cal W}_{(sp\pm c,[0])}(q,x)&=\frac{1}{\eta(q)} \sum_{n\in \mathbb{Z}} (\pm 1)^n q^{\frac{3}{2}\left(n+\frac{1}{2}\right)^2}\,,\\
\br^{\cal W}_{(sp\pm c,[2])}(q,x)&=\frac{1}{\eta(q)} \sum_{n\in \mathbb{Z}} (\pm 1)^{n+1} q^{\frac{3}{2}\left(n+\frac{1}{6}\right)^2}
(x^{n+\frac{1}{6}} \pm x^{-(n+\frac{1}{6})})\,.
\end{aligned}
\end{equation}
From these expressions we can read off the conformal weights of the
ground states in all $8$ sectors. Similarly, we can also determine the
maximal value $Q$ the U$(1)$ charge can assume among the ground states
of these sectors. In particular there are two sectors with $A$={\it
id}. The sector $(${\it id}$,[0])$ is the vacuum sector with $h=0$
and $Q=0$.

As discussed at the end of section 2.2, the branching functions associated
with the fixed points $(A,a_\ast)=(A,[2])$ can be decomposed into two
characters of our algebra ${\cal W}_2$. For $A =${\it id}  and {\it
v}, for example,
these characters read
\begin{equation}
\begin{aligned}\label{brv2}
\chi^{\cal W}_{(id,[2],1)} &=&  \frac{1}{\eta(q)} \sum_{n\in \mathbb{Z}}
q^{6 \left(n+\frac{1}{3}\right)^2} x^{2(n+\frac{1}{3})} \, ,  \\
\chi^{\cal W}_{(id,[2],2)}&=& \frac{1}{\eta(q)} \sum_{n\in \mathbb{Z}} q^{6\left(n+\frac{1}{3}\right)^2} x^{-2(n+\frac{1}{3})}\  ,
\\
\chi^{\cal W}_{(v,[2],1)} &=&  \frac{1}{\eta(q)} \sum_{n\in \mathbb{Z}}
q^{6 \left(n+\frac{1}{6}\right)^2} x^{2(n+\frac{1}{6})} \, ,   \\
\chi^{\cal W}_{(v,[2],2)}&=& \frac{1}{\eta(q)} \sum_{n\in \mathbb{Z}} q^{6\left(n+\frac{1}{6}\right)^2} x^{-2(n+\frac{1}{6})}\  .
\end{aligned}
\end{equation}
It is easy to check that these formulas agree with the expressions
\eqref{replacement1} and \eqref{replacement} when $x=1$.

Let us also display the necklace patterns for the two orbits $\{[0],[4]\}$
and $\{ [2]\}$. The affine Dynkin labels for these orbits are $[4,0]$ (or
$[0, 4]$) and $[2, 2]$. These correspond to the following two necklaces,
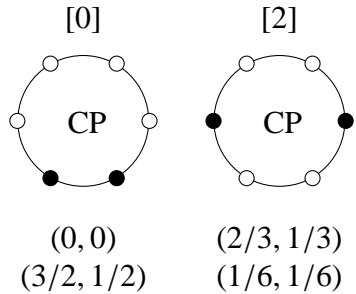
\begin{figure}[ht]
\begin{center}
\begin{minipage}{2.3cm}
\begin{center}[0]
\begin{picture}(56,80)
\necklacetwo{white}{ white}{ black}{ black}{ white}{ white}\put(22,43){CP}\end{picture}\vspace{-0.5cm}
$(0,0)$\\
$(3/2,1/2)$
\end{center}
\end{minipage}\hspace{0.2cm}
\begin{minipage}{2.3cm}
\begin{center}[2]
\begin{picture}(56,80)
\necklacetwo{white}{ black}{ white}{ white}{ black}{ white}\put(22,43){CP}\end{picture}\vspace{-0.5cm}
$(2/3,1/3)$\\
$(1/6,1/6)$
\end{center}
\end{minipage}\hspace{0.2cm}
\vspace{-0.5cm}
\end{center}
\end{figure}

\noindent In the two lines below the necklace we use tuples $(h,Q)$ to
display the ground state energy $h$ and maximal U(1) charge $Q$ among
the ground states of the sectors with $A$={\it id} (first line) and
$A$={\it v} (second line). In principle, there are also two sectors
with $A$={\it sp, c} which we do not show here. The label `CP' we
placed in the center of the two necklaces will be explained in the
next section.

Let us finally also spell out the full partition function of the
model. In the case at hand, our general expression
\eqref{partitionfct} reads as follows
\begin{align}\label{Z2resolved}
Z_2&=\sum_A \left(\vert \chi^{\cal W}_{(A,[0])}\vert^2 +
\vert \chi^{\cal W}_{(A,[2],1)} \vert^2 +\vert
\chi^{{\cal W}}_{(A,[2],2)}\vert^2   \right) \nn \\
&=\frac{1}{|\eta(q)|^2}\sum_{n,w\in\mathbb{Z}}q^{\frac{k_L^2}{2}}
\bar{q}^{\frac{k_R^2}{2}} x^{r k_L}
\bar{x}^{r k_R}  \quad\text{with}\quad k_{L,R}=\frac{n}{r}\pm\frac{w r}{2}
\,,\quad r=2\!\sqrt{3} \,
\end{align}
see appendix~\ref{appA3} for a few more details. The resummation
leading to the second line is in principle straight-forward. The
final result coincides with the usual partition function of a free
boson compactified on a circle of radius $2\!\sqrt{3}$.

\subsubsection*{$\mathbf{N=3}$}

After having gone through the example of \mbox{$N=2$} quite carefully,
we can now be a bit more sketchy with \mbox{$N=3$}. In this case we
obtain $|{\cal J}^0_3| =10$ representations of $\SU(3)_6$ with
vanishing monodromy charge. They can be grouped into four orbits,
three of which have length $N=3$ while the last one has length
$N_{[2,2]} = 1$. More explicitly, the orbits are given by
$\{\text{[0,0]},\text{[6,0]},\text{[0,6]}\}$,$\{\text{[1,1]},\text{[4,1]},
\text{[1,4]}\}$, $\{\text{[3,0]},\text{[0,3]},\text{[3,3]}\}$ and
$\{\text{[2,2]}\}$. The reader is invited to recover this list from
pairs of Young diagrams $Y'$ and $Y''$, as explained in section
\ref{integersec} above. The four orbits are associated with the following
four necklaces

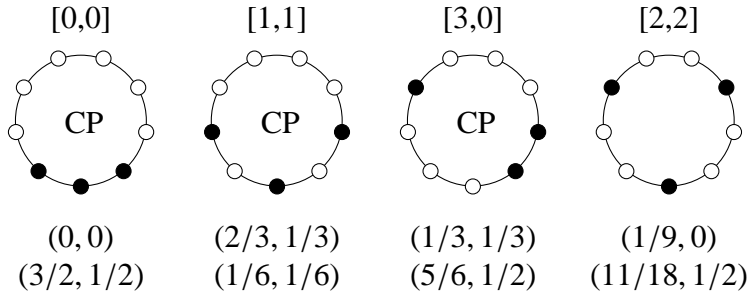
\begin{center}
\begin{minipage}{2.3cm}
\begin{center}[0,0]
\begin{picture}(56,80)
\necklacethree{white}{ white}{ white}{ black}{ black}{ black}{ white}{ white}{ white}\put(22,43){CP}\end{picture}\vspace{-0.5cm}
$(0,0)$\\
$(3/2,1/2)$
\end{center}
\end{minipage}\hspace{0.2cm}
\begin{minipage}{2.3cm}
\begin{center}[1,1]
\begin{picture}(56,80)
\necklacethree{white}{ white}{ black}{ white}{ black}{ white}{ black}{ white}{ white}\put(22,43){CP}\end{picture}\vspace{-0.5cm}
$(2/3,1/3)$\\
$(1/6,1/6)$
\end{center}
\end{minipage}\hspace{0.2cm}
\begin{minipage}{2.3cm}
\begin{center}[3,0]
\begin{picture}(56,80)
\necklacethree{white}{ black}{ white}{ white}{ white}{ black}{ black}{ white}{ white}\put(22,43){CP}\end{picture}\vspace{-0.5cm}
$(1/3,1/3)$\\
$(5/6,1/2)$
\end{center}
\end{minipage}\hspace{0.2cm}
\begin{minipage}{2.3cm}
\begin{center}[2,2]
\begin{picture}(56,80)
\necklacethree{white}{ black}{ white}{ white}{ black}{ white}{ white}{ black}{ white}\end{picture}\vspace{-0.5cm}
$(1/9,0)$\\
$(11/18,1/2)$
\end{center}
\end{minipage}\hspace{0.2cm}
\end{center}

The lines below these diagrams display again some information about
the associated branching functions $\br^{\cal W}_{(A,a)}$ for $A$={\it
id} and $A$={\it v}, namely the ground state energy $h$ and the
maximum $Q$ of the U(1) charge. These results can be read off from the
branching functions which we computed numerically, see
appendix~\ref{appB1} for the first few terms.

According to the general formula \eqref{partfu}, the function $\tilde Z_N$
takes the form
\begin{equation}
  \tilde Z_{3}(q) = \sum_A \left(\vert \br^{\cal W}_{(A,[0,0])}\vert^2
  + \vert \br^{\cal W}_{(A,[1,1])}\vert^2 +
   \vert \br^{\cal W}_{(A,[3,0])}\vert^2 +
  \textstyle\frac{1}{3}\vert \br^{\cal W}_{(A,[2,2])}\vert^2\right) \, .
\end{equation}
Once again, the $x$-dependent branching functions for the short orbit
can be decomposed into a sum of ${\cal W}_N$ characters. For instance,
the branching function\footnote{In comparison with appendix~\ref{appB1},
we have reintroduced the factor $q^{-c_3/{24}}$ in $\br^{\cal W}$ ($c_3=8/3$).}
\begin{align}
\br^{\cal W}_{(id, [2,2])}(q,x)
&= 1 + (2x^{4/3}+3x^{2/3}+5+3x^{-2/3}+2x^{-4/3}) \,q + O(q^2) \,,
\end{align}
can be written as the sum of three characters,
\begin{equation}
\begin{aligned}
\chi^{\cal W}_{(id,[2,2],1)}(q,x) &= \textstyle \frac{1}{2} \Big( \ch_{1/9}^{\NS, \rm ext}+ \widetilde \ch_{1/9}^{\NS,\rm ext}\Big)
= 1 + (x^{2/3}+3+x^{-2/3})\, q + O(q^2)  \,,\\[2mm]
\chi^{\cal W}_{(id,[2,2],p)}(q,x) &= \textstyle \frac{1}{2} \Big( \ch_{11/18}^{\NS,\rm ext}+ \widetilde \ch_{11/18}^{\NS,\rm ext}\Big)
=  (x^{4/3}+x^{2/3}+1+x^{-2/3}+x^{-4/3})\, q + O(q^2)
\end{aligned}
\end{equation}
for $p=2,3$. Here, the $\ch^{\NS,\rm ext}$ are  extended ${\cal N}=2$ characters,
as defined in \cite{Gopakumar}. After the resolution of the fixed point in the
sectors $(A, [2,2])$, we obtain the partition function
\begin{align}
Z_3(q,x)  = \sum_A \left(\vert \chi^{\cal W}_{(A,[0,0])}\vert^2
  + \vert \chi^{\cal W}_{(A,[1,1])}\vert^2 +
   \vert \chi^{\cal W}_{(A,[3,0])}\vert^2 +
  \vert \chi^{\cal W}_{(A,[2,2],1)}\vert^2 +  2   \vert \chi^{\cal W}_{(A,[2,2], 2)}\vert^2  \right) ,  \label{partfutN=3}
\end{align}
where all summands are considered as functions of both $q$ and $x$. Of course,
for $x=1$ we recover the expression \eqref{partitionfct} for the modular invariant
partition function we described above.

\subsubsection*{$\mathbf{N=4}$}

For $N=4$ there exist twelve different orbits which are labeled by the
following necklaces


\noindent{
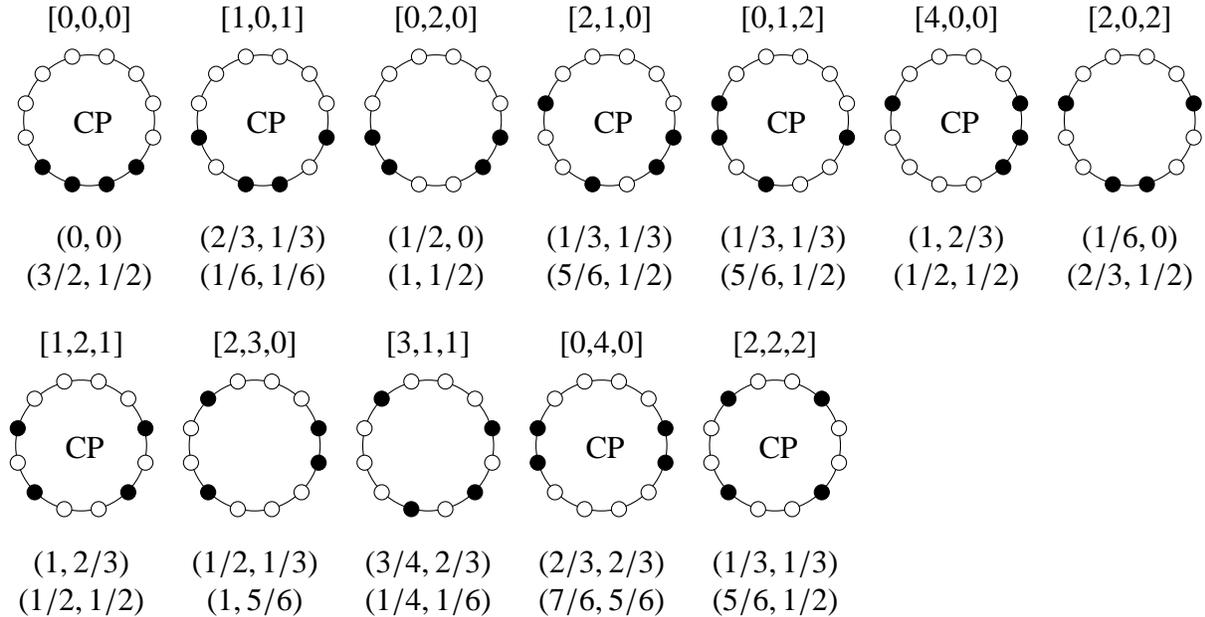
\begin{minipage}{2.cm}
\vspace{0.5cm}
\begin{center}[0,0,0]
\begin{picture}(56,80)
\necklacefoura{white}{ white}{ white}{ white}{ black}{ black}{ black}{ black}{ white}
\necklacefourb{white}{ white}{ white}\put(22,43){CP}\end{picture}\vspace{-0.5cm}
$(0,0)$\\
$(3/2,1/2)$
\end{center}
\end{minipage}\hspace{0.2cm}
\begin{minipage}{2.cm}
\vspace{0.5cm}
\begin{center}[1,0,1]
\begin{picture}(56,80)
\necklacefoura{white}{ white}{ white}{ black}{ white}{ black}{ black}{ white}{ black}
\necklacefourb{white}{ white}{ white}\put(22,43){CP}\end{picture}\vspace{-0.5cm}
$(2/3,1/3)$\\
$(1/6,1/6)$
\end{center}
\end{minipage}\hspace{0.2cm}
\begin{minipage}{2.cm}
\vspace{0.5cm}
\begin{center}[0,2,0]
\begin{picture}(56,80)
\necklacefoura{white}{ white}{ white}{ black}{ black}{ white}{ white}{ black}{ black}
\necklacefourb{white}{ white}{ white}\end{picture}\vspace{-0.5cm}
$(1/2,0)$\\
$(1,1/2)$
\end{center}
\end{minipage}\hspace{0.2cm}
\begin{minipage}{2.cm}
\vspace{0.5cm}
\begin{center}[2,1,0]
\begin{picture}(56,80)
\necklacefoura{white}{ white}{ black}{ white}{ white}{ black}{ white}{ black}{ black}
\necklacefourb{white}{ white}{ white}\put(22,43){CP}\end{picture}\vspace{-0.5cm}
$(1/3,1/3)$\\
$(5/6,1/2)$
\end{center}
\end{minipage}\hspace{0.2cm}
\begin{minipage}{2.cm}
\vspace{0.5cm}
\begin{center}[0,1,2]
\begin{picture}(56,80)
\necklacefoura{white}{ white}{ black}{ black}{ white}{ black}{ white}{ white}{ black}
\necklacefourb{white}{ white}{ white}\put(22,43){CP}\end{picture}\vspace{-0.5cm}
$(1/3,1/3)$\\
$(5/6,1/2)$
\end{center}
\end{minipage}\hspace{0.2cm}
\begin{minipage}{2.cm}
\vspace{0.5cm}
\begin{center}[4,0,0]
\begin{picture}(56,80)
\necklacefoura{white}{ white}{ black}{ white}{ white}{ white}{ white}{ black}{ black}
\necklacefourb{black}{ white}{ white}\put(22,43){CP}\end{picture}\vspace{-0.5cm}
$(1,2/3)$\\
$(1/2,1/2)$
\end{center}
\end{minipage}\hspace{0.2cm}
\begin{minipage}{2.cm}
\vspace{0.5cm}
\begin{center}[2,0,2]
\begin{picture}(56,80)
\necklacefoura{white}{ white}{ black}{ white}{ white}{ black}{ black}{ white}{ white}
\necklacefourb{black}{ white}{ white}\end{picture}\vspace{-0.5cm}
$(1/6,0)$\\
$(2/3,1/2)$
\end{center}
\end{minipage}\hspace{0.2cm}\\
\begin{minipage}{2.cm}
\vspace{0.5cm}
\begin{center}[1,2,1]
\begin{picture}(56,80)
\necklacefoura{white}{ white}{ black}{ white}{ black}{ white}{ white}{ black}{ white}
\necklacefourb{black}{ white}{ white}\put(22,43){CP}\end{picture}\vspace{-0.5cm}
$(1,2/3)$\\
$(1/2,1/2)$
\end{center}
\end{minipage}\hspace{0.2cm}
\begin{minipage}{2.cm}
\vspace{0.5cm}
\begin{center}[2,3,0]
\begin{picture}(56,80)
\necklacefoura{white}{ black}{ white}{ white}{ black}{ white}{ white}{ white}{ black}
\necklacefourb{black}{ white}{ white}\end{picture}\vspace{-0.5cm}
$(1/2,1/3)$\\
$(1,5/6)$
\end{center}
\end{minipage}\hspace{0.2cm}
\begin{minipage}{2.cm}
\vspace{0.5cm}
\begin{center}[3,1,1]
\begin{picture}(56,80)
\necklacefoura{white}{ black}{ white}{ white}{ white}{ black}{ white}{ black}{ white}
\necklacefourb{black}{ white}{ white}\end{picture}\vspace{-0.5cm}
$(3/4,2/3)$\\
$(1/4,1/6)$
\end{center}
\end{minipage}\hspace{0.2cm}
\begin{minipage}{2.cm}
\vspace{0.5cm}
\begin{center}[0,4,0]
\begin{picture}(56,80)
\necklacefoura{white}{ white}{ black}{ black}{ white}{ white}{ white}{ white}{ black}
\necklacefourb{black}{ white}{ white}\put(22,43){CP}\end{picture}\vspace{-0.5cm}
$(2/3,2/3)$\\
$(7/6,5/6)$
\end{center}
\end{minipage}\hspace{0.2cm}
\begin{minipage}{2.cm}
\vspace{0.5cm}
\begin{center}[2,2,2]
\begin{picture}(56,80)
\necklacefoura{white}{ black}{ white}{ white}{ black}{ white}{ white}{ black}{ white}
\necklacefourb{white}{ black}{ white}\put(22,43){CP}\end{picture}\vspace{-0.5cm}
$(1/3,1/3)$\\
$(5/6,1/2)$
\end{center}
\end{minipage}\hspace{0.2cm}}\\


Note that in this case there are two short orbits. While the orbit
$\{\text{[0,4,0]}, \text{[4,0,4]}\}$ has length $N_{[0,4,0]}=2$, the
element $a_\ast = \{\text{[2,2,2]}\}$ is fixed under the action of
$\gamma$ and hence gives an orbit of length $N_{[2,2,2]}=1$. The
branching functions of all orbits are displayed in appendix~\ref{appB2}.
Note that $4\times 2$ of these branching functions should be decomposed
into the sum of ${\cal W}_N$ characters since they are associated to
short orbits. For the remaining ones, the branching functions coincide
with the characters. We shall not discuss the resolution of fixed
points and the partition function of the system in any more detail.

\subsubsection*{$\mathbf{N=5}$}

Since $N=5$ is the first prime number beyond the minimal model
bound, the final case in our discussion is the most important
one.  For $N=5$ there are 41 different orbits which are labeled
by the following necklaces:


\noindent{
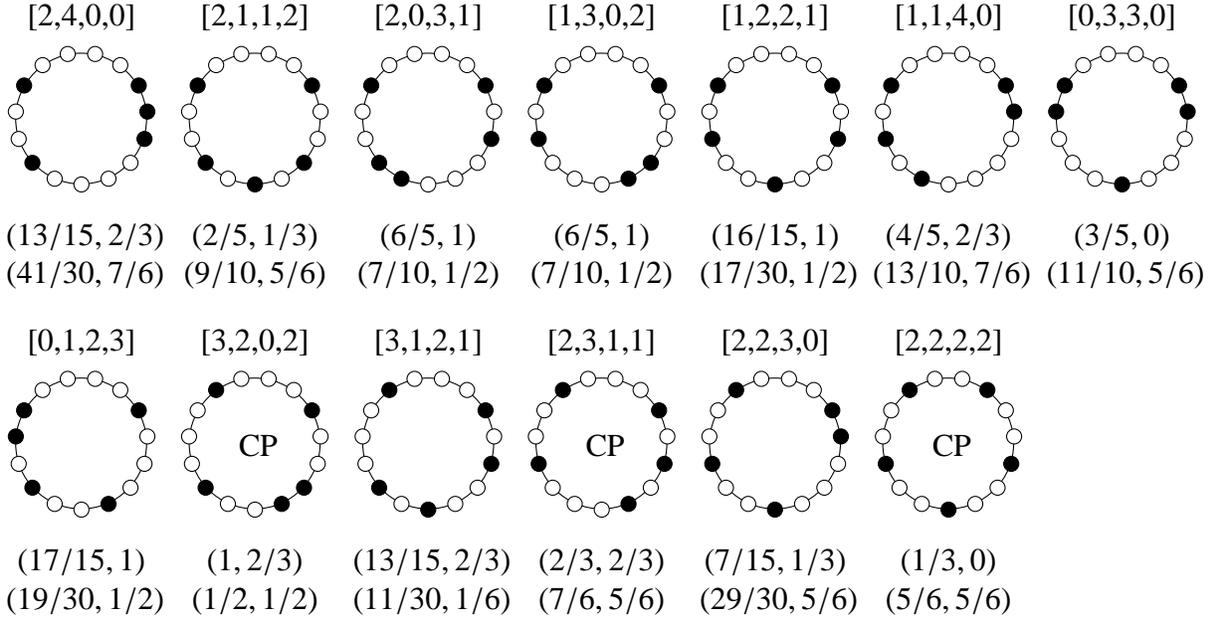
\begin{minipage}{2.cm}
\vspace{0.5cm}
\begin{center}[0,0,0,0]
\begin{picture}(56,80)
\necklacefivea{white}{ white}{ white}{ white}{ white}{ black}{ black}{ black}{ black}
\necklacefiveb{black}{ white}{ white}{ white}{ white}{ white}\put(22,43){CP}\end{picture}\vspace{-0.5cm}
$(0,0)$\\
$(3/2,1/2)$
\end{center}
\end{minipage}\hspace{0.2cm}
\begin{minipage}{2.cm}
\vspace{0.5cm}
\begin{center}[1,0,0,1]
\begin{picture}(56,80)
\necklacefivea{white}{ white}{ white}{ white}{ black}{ white}{ black}{ black}{ black}
\necklacefiveb{white}{ black}{ white}{ white}{ white}{ white}\put(22,43){CP}\end{picture}\vspace{-0.5cm}
$(2/3,1/3)$\\
$(1/6,1/6)$
\end{center}
\end{minipage}\hspace{0.2cm}
\begin{minipage}{2.cm}
\vspace{0.5cm}
\begin{center}[0,1,1,0]
\begin{picture}(56,80)
\necklacefivea{white}{ white}{ white}{ white}{ black}{ black}{ white}{ black}{ white}
\necklacefiveb{black}{ black}{ white}{ white}{ white}{ white}\end{picture}\vspace{-0.5cm}
$(7/15,0)$\\
$(29/30,1/2)$
\end{center}
\end{minipage}\hspace{0.2cm}
\begin{minipage}{2.cm}
\vspace{0.5cm}
\begin{center}[2,0,1,0]
\begin{picture}(56,80)
\necklacefivea{white}{ white}{ white}{ black}{ white}{ white}{ black}{ black}{ white}
\necklacefiveb{black}{ black}{ white}{ white}{ white}{ white}\put(22,43){CP}\end{picture}\vspace{-0.5cm}
$(1/3,1/3)$\\
$(5/6, 1/2)$
\end{center}
\end{minipage}\hspace{0.2cm}
\begin{minipage}{2.cm}
\vspace{0.5cm}
\begin{center}[1,2,0,0]
\begin{picture}(56,80)
\necklacefivea{white}{ white}{ white}{ black}{ white}{ black}{ white}{ white}{ black}
\necklacefiveb{black}{ black}{ white}{ white}{ white}{ white}\end{picture}\vspace{-0.5cm}
$(6/5,2/3)$\\
$(7/10,1/6)$
\end{center}
\end{minipage}\hspace{0.2cm}
\begin{minipage}{2.cm}
\vspace{0.5cm}
\begin{center}[0,1,0,2]
\begin{picture}(56,80)
\necklacefivea{white}{ white}{ white}{ white}{ black}{ black}{ white}{ black}{ black}
\necklacefiveb{white}{ white}{ black}{ white}{ white}{ white}\put(22,43){CP}\end{picture}\vspace{-0.5cm}
$(1/3,1/3)$\\
$(5/6, 1/2)$
\end{center}
\end{minipage}\hspace{0.2cm}
\begin{minipage}{2.cm}
\vspace{0.5cm}
\begin{center}[0,0,2,1]
\begin{picture}(56,80)
\necklacefivea{white}{ white}{ white}{ white}{ black}{ black}{ black}{ white}{ white}
\necklacefiveb{black}{ white}{ black}{ white}{ white}{ white}\end{picture}\vspace{-0.5cm}
$(6/5,2/3)$\\
$(7/10,1/6)$
\end{center}
\end{minipage}\hspace{0.2cm}\\
\begin{minipage}{2.cm}
\vspace{0.5cm}
\begin{center}[3,1,0,0]
\begin{picture}(56,80)
\necklacefivea{white}{ white}{ white}{ black}{ white}{ white}{ white}{ black}{ white}
\necklacefiveb{black}{ black}{ black}{ white}{ white}{ white}\put(22,43){CP}\end{picture}\vspace{-0.5cm}
$(1,2/3)$\\
$(1/2,1/2)$
\end{center}
\end{minipage}\hspace{0.2cm}
\begin{minipage}{2.cm}
\vspace{0.5cm}
\begin{center}[2,0,0,2]
\begin{picture}(56,80)
\necklacefivea{white}{ white}{ white}{ black}{ white}{ white}{ black}{ black}{ black}
\necklacefiveb{white}{ white}{ black}{ white}{ white}{ white}\end{picture}\vspace{-0.5cm}
$(1/5,0)$\\
$(7/10,1/2)$
\end{center}
\end{minipage}\hspace{0.2cm}
\begin{minipage}{2.cm}
\vspace{0.5cm}
\begin{center}[1,1,1,1]
\begin{picture}(56,80)
\necklacefivea{white}{ white}{ white}{ black}{ white}{ black}{ white}{ black}{ white}
\necklacefiveb{black}{ white}{ black}{ white}{ white}{ white}\put(22,43){CP}\end{picture}\vspace{-0.5cm}
$(1,2/3)$\\
$(1/2,1/2)$
\end{center}
\end{minipage}\hspace{0.2cm}
\begin{minipage}{2.cm}
\vspace{0.5cm}
\begin{center}[1,0,3,0]
\begin{picture}(56,80)
\necklacefivea{white}{ white}{ white}{ black}{ white}{ black}{ black}{ white}{ white}
\necklacefiveb{white}{ black}{ black}{ white}{ white}{ white}\end{picture}\vspace{-0.5cm}
$(4/5,1/3)$\\
$(13/10,5/6)$
\end{center}
\end{minipage}\hspace{0.2cm}
\begin{minipage}{2.cm}
\vspace{0.5cm}
\begin{center}[0,3,0,1]
\begin{picture}(56,80)
\necklacefivea{white}{ white}{ white}{ black}{ black}{ white}{ white}{ white}{ black}
\necklacefiveb{black}{ white}{ black}{ white}{ white}{ white}\end{picture}\vspace{-0.5cm}
$(4/5,1/3)$\\
$(13/10,5/6)$
\end{center}
\end{minipage}\hspace{0.2cm}
\begin{minipage}{2.cm}
\vspace{0.5cm}
\begin{center}[0,2,2,0]
\begin{picture}(56,80)
\necklacefivea{white}{ white}{ white}{ black}{ black}{ white}{ white}{ black}{ white}
\necklacefiveb{white}{ black}{ black}{ white}{ white}{ white}\put(22,43){CP}\end{picture}\vspace{-0.5cm}
$(2/3,2/3)$\\

$(7/6,5/6)$
\end{center}
\end{minipage}\hspace{0.2cm}
\begin{minipage}{2.cm}
\vspace{0.5cm}
\begin{center}[0,0,1,3]
\begin{picture}(56,80)
\necklacefivea{white}{ white}{ white}{ black}{ black}{ black}{ white}{ black}{ white}
\necklacefiveb{white}{ white}{ black}{ white}{ white}{ white}\put(22,43){CP}\end{picture}\vspace{-0.5cm}
$(1,2/3)$\\
$(1/2,1/2)$
\end{center}
\end{minipage}\hspace{0.2cm}\\
\begin{minipage}{2.cm}
\vspace{0.5cm}
\begin{center}[5,0,0,0]
\begin{picture}(56,80)
\necklacefivea{white}{ white}{ black}{ white}{ white}{ white}{ white}{ white}{ black}
\necklacefiveb{black}{ black}{ black}{ white}{ white}{ white}\put(22,43){CP}\end{picture}\vspace{-0.5cm}
$(2/3,2/3)$\\
$(7/6,5/6)$
\end{center}
\end{minipage}\hspace{0.2cm}
\begin{minipage}{2.cm}
\vspace{0.5cm}
\begin{center}[3,0,1,1]
\begin{picture}(56,80)
\necklacefivea{white}{ white}{ black}{ white}{ white}{ white}{ black}{ black}{ white}
\necklacefiveb{black}{ white}{ black}{ white}{ white}{ white}\end{picture}\vspace{-0.5cm}
$(4/5,2/3)$\\
$(3/10,1/6)$
\end{center}
\end{minipage}\hspace{0.2cm}
\begin{minipage}{2.cm}
\vspace{0.5cm}
\begin{center}[2,2,0,1]
\begin{picture}(56,80)
\necklacefivea{white}{ white}{ black}{ white}{ white}{ black}{ white}{ white}{ black}
\necklacefiveb{black}{ white}{ black}{ white}{ white}{ white}\put(22,43){CP}\end{picture}\vspace{-0.5cm}
$(2/3,2/3)$\\
$(7/6,5/6)$
\end{center}
\end{minipage}\hspace{0.2cm}
\begin{minipage}{2.cm}
\vspace{0.5cm}
\begin{center}[2,1,2,0]
\begin{picture}(56,80)
\necklacefivea{white}{ white}{ black}{ white}{ white}{ black}{ white}{ black}{ white}
\necklacefiveb{white}{ black}{ black}{ white}{ white}{ white}\end{picture}\vspace{-0.5cm}
$(8/15,1/3)$\\
$(31/30,5/6)$
\end{center}
\end{minipage}\hspace{0.2cm}
\begin{minipage}{2.cm}
\vspace{0.5cm}
\begin{center}[1,3,1,0]
\begin{picture}(56,80)
\necklacefivea{white}{ white}{ black}{ white}{ black}{ white}{ white}{ white}{ black}
\necklacefiveb{white}{ black}{ black}{ white}{ white}{ white}\put(22,43){CP}\end{picture}\vspace{-0.5cm}
$(4/3,1)$\\
$(5/6,5/6)$
\end{center}
\end{minipage}\hspace{0.2cm}
\begin{minipage}{2.cm}
\vspace{0.5cm}
\begin{center}[1,1,0,3]
\begin{picture}(56,80)
\necklacefivea{white}{ white}{ white}{ black}{ white}{ black}{ white}{ black}{ black}
\necklacefiveb{white}{ white}{ white}{ black}{ white}{ white}\end{picture}\vspace{-0.5cm}
$(4/5,2/3)$\\
$(3/10,1/6)$
\end{center}
\end{minipage}\hspace{0.2cm}
\begin{minipage}{2.cm}
\vspace{0.5cm}
\begin{center}[1,0,2,2]
\begin{picture}(56,80)
\necklacefivea{white}{ white}{ white}{ black}{ white}{ black}{ black}{ white}{ white}
\necklacefiveb{black}{ white}{ white}{ black}{ white}{ white}\put(22,43){CP}\end{picture}\vspace{-0.5cm}
$(2/3,2/3)$\\
$(7/6,5/6)$
\end{center}
\end{minipage}\hspace{0.2cm}\\
\begin{minipage}{2.cm}
\vspace{0.5cm}
\begin{center}[0,5,0,0]
\begin{picture}(56,80)
\necklacefivea{white}{ white}{ black}{ black}{ white}{ white}{ white}{ white}{ white}
\necklacefiveb{black}{ black}{ black}{ white}{ white}{ white}\put(22,43){CP}\end{picture}\vspace{-0.5cm}
$(1,1)$\\
$(3/2,7/6)$
\end{center}
\end{minipage}\hspace{0.2cm}
\begin{minipage}{2.cm}
\vspace{0.5cm}
\begin{center}[0,2,1,2]
\begin{picture}(56,80)
\necklacefivea{white}{ white}{ white}{ black}{ black}{ white}{ white}{ black}{ white}
\necklacefiveb{black}{ white}{ white}{ black}{ white}{ white}\end{picture}\vspace{-0.5cm}
$(8/15,1/3)$\\
$(31/30,5/6)$
\end{center}
\end{minipage}\hspace{0.2cm}
\begin{minipage}{2.cm}
\vspace{0.5cm}
\begin{center}[0,1,3,1]
\begin{picture}(56,80)
\necklacefivea{white}{ white}{ white}{ black}{ black}{ white}{ black}{ white}{ white}
\necklacefiveb{white}{ black}{ white}{ black}{ white}{ white}\put(22,43){CP}\end{picture}\vspace{-0.5cm}
$(4/3,1)$\\
$(5/6,5/6)$
\end{center}
\end{minipage}\hspace{0.2cm}
\begin{minipage}{2.cm}
\vspace{0.5cm}
\begin{center}[4,1,0,1]
\begin{picture}(56,80)
\necklacefivea{white}{ white}{ black}{ white}{ white}{ white}{ white}{ black}{ white}
\necklacefiveb{black}{ black}{ white}{ black}{ white}{ white}\end{picture}\vspace{-0.5cm}
$(2/5,1/3)$\\
$(9/10,5/6)$
\end{center}
\end{minipage}\hspace{0.2cm}
\begin{minipage}{2.cm}
\vspace{0.5cm}
\begin{center}[4,0,2,0]
\begin{picture}(56,80)
\necklacefivea{white}{ white}{ black}{ white}{ white}{ white}{ white}{ black}{ black}
\necklacefiveb{white}{ white}{ black}{ black}{ white}{ white}\end{picture}\vspace{-0.5cm}
$(4/15,0)$\\
$(23/30,1/2)$
\end{center}
\end{minipage}\hspace{0.2cm}
\begin{minipage}{2.cm}
\vspace{0.5cm}
\begin{center}[3,2,1,0]
\begin{picture}(56,80)
\necklacefivea{white}{ white}{ black}{ white}{ white}{ white}{ black}{ white}{ white}
\necklacefiveb{black}{ white}{ black}{ black}{ white}{ white}\end{picture}\vspace{-0.5cm}
$(17/15,1)$\\
$(19/30,1/2)$
\end{center}
\end{minipage}\hspace{0.2cm}
\begin{minipage}{2.cm}
\vspace{0.5cm}
\begin{center}[3,0,0,3]
\begin{picture}(56,80)
\necklacefivea{white}{ white}{ black}{ white}{ white}{ white}{ black}{ black}{ black}
\necklacefiveb{white}{ white}{ white}{ black}{ white}{ white}\end{picture}\vspace{-0.5cm}
$(3/5,1/3)$\\
$(11/10,5/6)$
\end{center}
\end{minipage}\hspace{0.2cm}\\
\begin{minipage}{2.cm}
\vspace{0.5cm}
\begin{center}[2,4,0,0]
\begin{picture}(56,80)
\necklacefivea{white}{ white}{ black}{ white}{ white}{ black}{ white}{ white}{ white}
\necklacefiveb{white}{ black}{ black}{ black}{ white}{ white}\end{picture}\vspace{-0.5cm}
$(13/15,2/3)$\\
$(41/30,7/6)$
\end{center}
\end{minipage}\hspace{0.2cm}
\begin{minipage}{2.cm}
\vspace{0.5cm}
\begin{center}[2,1,1,2]
\begin{picture}(56,80)
\necklacefivea{white}{ white}{ black}{ white}{ white}{ black}{ white}{ black}{ white}
\necklacefiveb{black}{ white}{ white}{ black}{ white}{ white}\end{picture}\vspace{-0.5cm}
$(2/5,1/3)$\\
$(9/10,5/6)$
\end{center}
\end{minipage}\hspace{0.2cm}
\begin{minipage}{2.cm}
\vspace{0.5cm}
\begin{center}[2,0,3,1]
\begin{picture}(56,80)
\necklacefivea{white}{ white}{ black}{ white}{ white}{ black}{ black}{ white}{ white}
\necklacefiveb{white}{ black}{ white}{ black}{ white}{ white}\end{picture}\vspace{-0.5cm}
$(6/5,1)$\\
$(7/10,1/2)$
\end{center}
\end{minipage}\hspace{0.2cm}
\begin{minipage}{2.cm}
\vspace{0.5cm}
\begin{center}[1,3,0,2]
\begin{picture}(56,80)
\necklacefivea{white}{ white}{ black}{ white}{ black}{ white}{ white}{ white}{ black}
\necklacefiveb{black}{ white}{ white}{ black}{ white}{ white}\end{picture}\vspace{-0.5cm}
$(6/5,1)$\\
$(7/10,1/2)$
\end{center}
\end{minipage}\hspace{0.2cm}
\begin{minipage}{2.cm}
\vspace{0.5cm}
\begin{center}[1,2,2,1]
\begin{picture}(56,80)
\necklacefivea{white}{ white}{ black}{ white}{ black}{ white}{ white}{ black}{ white}
\necklacefiveb{white}{ black}{ white}{ black}{ white}{ white}\end{picture}\vspace{-0.5cm}
$(16/15,1)$\\
$(17/30,1/2)$
\end{center}
\end{minipage}\hspace{0.2cm}
\begin{minipage}{2.cm}
\vspace{0.5cm}
\begin{center}[1,1,4,0]
\begin{picture}(56,80)
\necklacefivea{white}{ white}{ black}{ white}{ black}{ white}{ black}{ white}{ white}
\necklacefiveb{white}{ white}{ black}{ black}{ white}{ white}\end{picture}\vspace{-0.5cm}
$(4/5,2/3)$\\
$(13/10,7/6)$
\end{center}
\end{minipage}\hspace{0.2cm}
\begin{minipage}{2.cm}
\vspace{0.5cm}
\begin{center}[0,3,3,0]
\begin{picture}(56,80)
\necklacefivea{white}{ white}{ black}{ black}{ white}{ white}{ white}{ black}{ white}
\necklacefiveb{white}{ white}{ black}{ black}{ white}{ white}\end{picture}\vspace{-0.5cm}
$(3/5,0)$\\
$(11/10,5/6)$
\end{center}
\end{minipage}\hspace{0.2cm}\\
\begin{minipage}{2.cm}
\vspace{0.5cm}
\begin{center}[0,1,2,3]
\begin{picture}(56,80)
\necklacefivea{white}{ white}{ black}{ black}{ white}{ black}{ white}{ white}{ black}
\necklacefiveb{white}{ white}{ white}{ black}{ white}{ white}\end{picture}\vspace{-0.5cm}
$(17/15,1)$\\
$(19/30,1/2)$
\end{center}
\end{minipage}\hspace{0.2cm}
\begin{minipage}{2.cm}
\vspace{0.5cm}
\begin{center}[3,2,0,2]
\begin{picture}(56,80)
\necklacefivea{white}{ black}{ white}{ white}{ white}{ black}{ white}{ white}{ black}
\necklacefiveb{black}{ white}{ white}{ black}{ white}{ white}\put(22,43){CP}\end{picture}\vspace{-0.5cm}
$(1,2/3)$\\
$(1/2,1/2)$
\end{center}
\end{minipage}\hspace{0.2cm}
\begin{minipage}{2.cm}
\vspace{0.5cm}
\begin{center}[3,1,2,1]
\begin{picture}(56,80)
\necklacefivea{white}{ black}{ white}{ white}{ white}{ black}{ white}{ black}{ white}
\necklacefiveb{white}{ black}{ white}{ black}{ white}{ white}\end{picture}\vspace{-0.5cm}
$(13/15,2/3)$\\
$(11/30,1/6)$
\end{center}
\end{minipage}\hspace{0.2cm}
\begin{minipage}{2.cm}
\vspace{0.5cm}
\begin{center}[2,3,1,1]
\begin{picture}(56,80)
\necklacefivea{white}{ black}{ white}{ white}{ black}{ white}{ white}{ white}{ black}
\necklacefiveb{white}{ black}{ white}{ black}{ white}{ white}\put(22,43){CP}\end{picture}\vspace{-0.5cm}
$(2/3,2/3)$\\
$(7/6,5/6)$
\end{center}
\end{minipage}\hspace{0.2cm}
\begin{minipage}{2.cm}
\vspace{0.5cm}
\begin{center}[2,2,3,0]
\begin{picture}(56,80)
\necklacefivea{white}{ black}{ white}{ white}{ black}{ white}{ white}{ black}{ white}
\necklacefiveb{white}{ white}{ black}{ black}{ white}{ white}\end{picture}\vspace{-0.5cm}
$(7/15,1/3)$\\
$(29/30,5/6)$
\end{center}
\end{minipage}\hspace{0.2cm}
\begin{minipage}{2.cm}
\vspace{0.5cm}
\begin{center}[2,2,2,2]
\begin{picture}(56,80)
\necklacefivea{white}{ black}{ white}{ white}{ black}{ white}{ white}{ black}{ white}
\necklacefiveb{white}{ black}{ white}{ white}{ black}{ white}\put(22,43){CP}\end{picture}\vspace{-0.5cm}
$(1/3,0)$\\
$(5/6,5/6)$
\end{center}
\end{minipage}\vspace{0.5cm}\hspace{0.2cm} }    \\
As usual, the length of the orbit $a_\ast = \{\text{[2,2,2,2]}\}$ is
$N_{a_\ast}=1$. All other orbits are of maximal length $N$. The first
few terms of the branching functions are displayed in appendix~\ref{appB3}.

Let us briefly describe how to resolve the fixed point when we work
with $x$-dependent branching functions and characters. One may find
the following expression for the branching function
\begin{align}
\br^{\cal W}_{(id, [2,2,2,2])}
&= 1 +(4y^2 + 19 y^{4/3}+ 36 y^{2/3}+47) \,q + O(q^2) \,,
\end{align}
in appendix~\ref{appB3}. It can be written as a sum of five functions,
\begin{equation}
\begin{aligned}
\chi^{\cal W}_{(id,[2,2,2,2],1)} &=
1 + (3 y^{4/3}+ 8 y^{2/3}+11)\, q + O(q^2)  \,, \\[2mm]
\chi^{\cal W}_{(id,[2,2,2,2], p)} &=
 (y^2 + 4 y^{4/3}+ 7 y^{2/3}+9)\, q + O(q^2)  \qquad
\end{aligned}
\end{equation}
for $p=2,...,5$, which we propose for the characters. Here we have introduced the shorthand $y^n\equiv
x^n+x^{-n}$. Note that for $x=1$ the coefficients of $q$ in the
characters must equal $165/5=33$. Then, after resolution of the
fixed point, we get
\begin{align} \label{partfuN=5}
 Z_5 =  \sum_A \Big( \sum_{\{\a\} \in {\cal O}_5/\{[2,2,2,2]\}}  \vert
    \chi^{\cal W}_{(A,a)}\vert^2 +  | \chi^{\cal W}_{(A, [2,2,2,2],1)}|^2 + 4 | \chi^{\cal W}_{(A, [2,2,2,2],2)}|^2 \Big)\,.
\end{align}
This concludes our brief discussion of branching functions, characters
and partition functions for the examples with $N \leq 5$.

\setcounter{equation}{0}
\section{Chiral primary fields}

In this section we will describe the main results of this work. We
have described the chiral symmetry ${\cal W}_N$ and the complete
modular invariant partition function $Z_N$ for a family of field
theories with ${\cal N}=(2,2)$ superconformal symmetry.  Our goal now is
to determine the chiral primaries of these models.  Since we know
how the spectrum of the model is built from the various
representations of the chiral algebra ${\cal W}_N$ all that is
left to do is to find (anti-)chiral primaries in the individual
sectors. In principle this is straightforward once the characters of
the chiral algebra are known. Indeed, for $N\leq 5$, the chiral
primaries can be read off from the $q$-expanded branching functions
listed in appendix~\ref{appB}.  In section~\ref{secbound}, we show
that there exists an upper bound on the conformal weight of a chiral
primary. In order to organize the chiral primaries, we will then
define and discuss in section~\ref{secregular} the class of {\em
regular} chiral primaries. In section~\ref{secexample}, we discuss a
few examples and show that for $N \leq 3$ there are no other chiral
primaries besides the regular ones. This will change for theories
with $N \geq 4$, as we shall show in section~\ref{secexceptional}.

\subsection{Bound on the dimension of chiral primaries }\label{secbound}

There are a few general results on the dimension of chiral primaries
that are useful to discuss before we get into concrete examples. In
any ${\cal N}=2$ superconformal field theory, the conformal weight of
chiral primaries is bounded from above by
\begin{equation}\label{cbound}
h(\phi_\cp) \leq \frac{c}{6}
\end{equation}
where $c$ is the central charge of the Virasoro algebra~\cite{Lerche}. This bound
is independent of the sector in which the chiral primary resides.

In order to derive stronger sector dependent bounds, we recall that
the fields in the numerator theory satisfy $h^\N \geq 3 |Q^\N|$.
States that make it into the coset sector $(A,\{a\})$ contain
the highest weight vector of a $\SU(N)_{2N}$ representation \mbox{$b \in
\{a\}$} in the orbit of $a$. The latter has weight $h^\D_b$ and charge
$Q^\D_b =0$. For the dimension $h$ and  charge $Q$ of the coset
fields we obtain the constraint $h + h^\D_b = h^\N \geq 3 |Q^\N| =
3|Q|$ and consequently for coset states $\phi$ in the sector
$(A,\{a\})$,
$$ h(\phi) \geq 3|Q(\phi)| - \mbox{\it min}_{b\in \{a\}}
(h^\D_b)\ . $$
For (anti-)chiral primaries $\phi_{\cp}$ with $h(\phi_{\cp}) =
|Q(\phi_{\cp})|$ this inequality implies that
\begin{align}
2h(\phi_{\cp}) \leq  \mbox{\it min}_{b\in \{a\}} (h^\D_b) \quad
\mbox{ or } \quad
h(\phi_{\cp}) \leq \mbox{\it min}_{b\in \{a\}} \left(\frac{C_2(b)}{6N}
\right) \ . \label{bound}
\end{align}
In addition to this constraint, the $\U(1)$ charges must also satisfy
$|Q|= k/6$ ($k \in \mathbb{N}_0$). It is easy to see that this implies
$$\mbox{\it min}_{b\in
\{a\}} \left(\frac{C_2(b)}{N}\right) \in \mathbb{Z} $$ if the sector
$(A,\{a\})$ is to contain a chiral primary and $N$ is odd.  For even
$N$ a similar condition holds with $N$ replaced by $N/2$.

As we shall see below there exist some important sectors for which
this bound is so strong that it does not permit chiral primaries
above the ground states. In other sectors, however, our bound
\eqref{bound} is much less powerful. This applies in particular
to those that are associated with the fixed point $a_\ast$. In
fact, in the representation $a_\ast$, the quadratic Casimir
assumes its largest eigenvalue
$$ C_2(a_\ast) = \frac{1}{3} N (N^2 -1) \quad \mbox{ or }
\quad \frac{C_2(a_\ast)}{6N} = \frac{c_N}{6}\ . $$
Hence, in the fixed point sectors our bound \eqref{bound} coincides
with the universal bound \eqref{cbound}. This appears to leave a lot
of room for chiral primaries.

\subsection{Regular chiral primaries}\label{secregular}

As we stressed before, there exists a large set of chiral
primaries that may be constructed very explicitly for any
value of $N$. Their description is particularly simple when
we use our parametrization of orbits in terms of two Young
diagrams $Y'$ and $Y''$, see section 3.1. We will determine
${\cal W}$ sectors containing regular chiral primaries in
the first subsection and then count regular chiral primaries
of the full (non-chiral) conformal field theory in the second.

\subsubsection{Parametrization and properties}

In section 3 we constructed all solutions $a \in {\cal J}_N$ of
the vanishing monodromy condition $Q(a) =0$ in terms of a pair of
Young diagrams $Y' = Y'(a)$ and $Y'' = Y''(a)$. As indicated in
our notation we now think of these Young diagrams as functions of
the sector label $a$. This map is obtained by reversing our formula
\eqref{generation} for the construction of $Y(a)$ from $Y'$ and
$Y''$. As we shall show below, for elements of the following subset
\begin{equation}
\S_N = \{ a=(l_1,\dots,l_{r})|  Y'(a) = Y''(a); Q(a) = 0 \} \subset
{\cal J}^0_N
\end{equation}
we can find a regular chiral primary in the coset sectors $(A,\{a\})$
with $A =${\it id} if $|Y'|$ is even and $A = v$ otherwise.

\begin{figure}
\hspace{4cm}
\begin{picture}(60,30)
 \put(0,20) { \yng(2,1)}
 \put(14.5,45) { \rotatebox{180}{{\color{darkgreen} \yng(3,2)}}}
  \put(100,30) {$\longrightarrow$   }
 \put(160,10)  {\color{darkgreen} \yng(2,2,1)}
 \put(185,23)  { \yng(2,1)}
\end{picture}
\caption{Generation of a representation $Y \in \S_4$. The  Young diagram
 $Y'$ with lengths $(l'_1, l'_2)=(2,1)$ (black) induces the $\SU(4)$
 representation $Y$ with lengths $(l_1, l_2,
 l_3)=(4,3,1)$.}\label{figYT}
\end{figure}
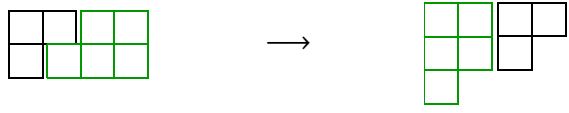

There is a relatively simple geometrical construction of the
Young diagrams $Y(a)$ that can be obtained with $Y' = Y''$. In
fact, it follows from eq.\ (\ref{generation}) that a Young diagram
$Y(a), a \in \S_N$ is obtained from $Y'$ by first completing $Y'$
to an $r' \times N$ rectangular Young diagram and then attaching the
(rotated) `complementary' diagram $(N-l'_{r'}, ..., N-l'_{1})$ from
the left to the original Young diagram $Y'$. An example is shown in
figure~\ref{figYT}.

Note that the condition $Y'(a) = Y''(a)$ is not invariant under the
action of the identification current. So, in order to find out whether
a sector $(A,\{a\})$ contains a regular chiral primary, one has to
check the condition $Y'(b) = Y''(b)$ for all $b \in \{a\}$.

There are two important remarks we have to make concerning the precise
relation between coset sectors containing regular chiral primaries and
elements of the set $\S_N$. The first one concerns the sectors obtained
from $a_\ast = [2,2,\dots,2]$ that give rise to the unique fixed points
for $N$ prime. A moment of thought about the construction we sketched
above shows that $a_\ast$ can never be in $\S_N$ unless $N=2$. Hence
the distinction between coset sectors and labels $(A,\{a\})$, as well
as all our discussion of fixed point resolutions, is not relevant for
the discussion of regular chiral primaries when $N$ is a prime number.

More importantly, we want to stress that there exist orbits $\{a\}
\in {\cal O}_N$ that contain two elements from $\S_N$. It is not too
difficult to list these orbits explicitly. From the general
construction of $\S_N$ we can infer that
$$ a_\nu = [0,...,0,N,0,...,0] \in \S_N $$
for $\nu = 1, \dots, N-1$. Here, the only non-zero entry $N$ can appear
in any position $\nu$, i.e. $\lambda_\nu = N$. Field identifications can
map $a_\nu$ to $a_{N-\nu}$ so that we have now found  $\left\lfloor(N-1)
/2\right\rfloor$ orbits that contain two elements of $\S_N$. One may
also argue that these are the only ones that contain more than one
element of $\S_N$.

For coset sectors $(A,\{a\})$ with $a \in \S_N$ there exists a simple
formula to compute their exact conformal weight. By eq.\ (\ref{C2generalized})
the quadratic Casimir of a representation $a \in \S_N$ is simply
$C_2(a)=n' N$. This implies that the conformal weight of the
ground states is
\begin{align}
h(\psi_{(A,\{a\})}) &=
 \frac{C_2(a)}{6N} = \frac{n'}{6} \,,\label{cw}
\end{align}
for $a\in \S_N$. Let us add that the number $n' = |Y'| = |Y''|$ of boxes
in $Y'$ is given in terms of the representation labels $(l_1, ..., l_r)$
of $a$ by
\begin{align}
n' = \sum_{i=1}^{r'} l'_i = \sum_{i=1}^{r'} (l_i - r')  \label{ra}
\end{align}
with $r'=n/N$ and where $n=\sum_{i=1}^r l_i$ is the number of boxes of
$Y = Y(a)$. With the help or sector dependent bound from the previous
subsection, see eq.\ \eqref{bound}, we can now show that in all the
sectors associated with $a \in \S_N$, chiral primaries must be
ground states of the ${\cal W}$ algebra. In fact, by combining the
conformal weight (\ref{cw}) with the bound (\ref{bound}), we find
$$ \frac{n'}{6} = h(\psi_{(A,\{a\})}) \leq h(\phi_{\cp}) \leq
   \mbox{\it min}_{b\in \{a\}} \left(\frac{C_2(b)}{6N}\right) =
   \frac{n'}{6} \ .
$$
Hence, these sectors cannot contain any chiral primaries in addition
to the ones we will find among their ground states.

\subsubsection{Counting of regular chiral primaries}

Before we look into examples let us count the regular chiral
primaries along with their conformal weight. This will proceed
in several steps. First we shall count the number of elements
in $\S_N$, then we employ the result to count the number
representations of our chiral algebra ${\cal W}$ that contain
a regular chiral primary and finally we determine the
counting function for regular chiral primaries from the
full partition function of the model, at least for $N$ prime.

Our description of the set $\S_N$ in terms of Young diagrams
$Y' = Y''$ makes it an easy task to determine $|\S_N|$. The
conditions for the choice of $Y'$ and $Y''$ we spelled out
before eq.\ \eqref{generation}. They imply that diagrams $Y'$ 
corresponding to elements in $\S_N$ must fit into a rectangle of
size $r' \times c'$ with $r'+c'=N$.  Such Young diagrams are
counted through the series
\begin{align}\label{tTN}
\tilde T_N(\q) =  \sum_{k=0}^N
\begin{bmatrix} N \\ k \end{bmatrix}_\q
- \sum_{k=0}^{N-1}
 \begin{bmatrix} N-1 \\ k \end{bmatrix}_\q
=  \sum_{k=0}^{N-1} \q^k
\begin{bmatrix} N-1 \\ k \end{bmatrix}_\q\,,
\end{align}
which is denoted by A161161 in \cite{OEIS}. The $\q$-binomial coefficient
that multiplies $\q^k$ counts all Young diagrams $Y'$ that fit into a
rectangle with $(N-1-k) \times k$ boxes. The factor $\q^k$ corresponds
to attaching to each of these Young diagrams from the left a single
column of $k$ boxes.  As a consequence, the individual summands in
eq.\ (\ref{tTN}) count the number of Young diagrams fitting into a
rectangle with $(N-k)\times k$ boxes. By the binomial theorem we find
$$| \S_N |= \tilde T_N(1)=\sum_{k=0}^{N-1}\scalebox{0.7}{$\begin{pmatrix} N-1 \\ k
\end{pmatrix}$}=2^{N-1}\ . $$
Let us list also the coefficients $\tilde t^N_n$ of the function $\tilde T_N(\q)=
\sum_n \tilde t^N_n \q^n$ for all values with $N \leq 7$,
\begin{align}
N=2:\quad &1, 1\nn\\
N=3:\quad &1, 1, 2\nn\\
N=4:\quad &1, 1, 2, 3, 1\nn\\
N=5:\quad &1, 1, 2, 3, 5, 2, 2\nn\\
N=6:\quad &1, 1, 2, 3, 5, 7, 5,  4,  3,  1\nn\\
N=7:\quad &1, 1, 2, 3, 5, 7, 11, 8,  9,  7,  6,  2,  2\,. \nn
\end{align}
Let us note in passing that, at large $N$, the coefficients $\tilde t^N_n$ of
$\tilde T_N(\q)$ coincide with the number $p_n$ of partitions of $n$, i.e.\
\begin{align} \label{TNlimit}
 \lim_{N\rightarrow \infty} \tilde T_N(\q) = \prod_{k=1}^{\infty} \frac{1}{
 1-\q^k} = \sum_{n=0}^{\infty} p(n)\q^n\,.
\end{align}
In order to count the number of ${\cal W}$ representations that
contain a regular chiral primary we recall two facts discussed
above. The first one concerns the fixed point resolution. When
$N>2$ is prime, there is only one short orbit $a_\ast$ and since
$a_\ast$ is not a element of $\S_N$ the counting of {\em regular}
chiral primaries is not affected by the fixed point resolution.
On the other hand, there are a few orbits that contain two
elements of $\S_N$. These need to be subtracted from the
counting function $\tilde T_N(\q)$ in order to obtain a counting
function $T_N(\q)$ for ${\cal W}$ sectors containing regular
chiral primaries
\begin{equation}  \label{TN}
 T_N(\q) :=  \tilde T_N (\q)
- \sum_{n=1}^{\left\lfloor\!\frac{N-1}{2}\!\right\rfloor}
\q^{n(N-n)/6} \,.
\end{equation}
As explained above, the over-counting we are trying make up for
is associated with the Dynkin labels $a_\nu \in \S_N$. Since the
sector $(A,a_\nu)$ containing the associated regular chiral
primary has conformal weight $h_{(A,[0...N...0])} = n(N-n)/6$,
we have included the appropriate power of $\q$ in our subtraction.

After this preparation we can now turn to the counting of (regular)
chiral primaries. By their very definition, (anti-)chiral primaries
are fields in the Neveu-Schwarz sector of the theory for which the
conformal weight $h$ and the U(1) charge $Q$ satisfy  $h= \pm Q$.
Here, the upper sign applies to chiral primaries while the lower one
is relevant for anti-chiral primaries.  It is then obvious that
chiral primaries are counted by
\begin{align}
  Z_{N}^{\cp}(q,\bar q) &=
  \frac{1}{(2\pi i)^2} \oint \frac{dx}{x}
   \oint \frac{d\bar x}{\bar x} \ Z^{\text{NS}}_N(qx^{-2},
   \bar q\bar x^{-2}, x,\bar x)
\end{align}
where $Z^{\text NS}_N$ denotes the contribution from the NS sector
of the model, i.e. the summands $A$=$id$ and $A=v$, to the full
(resolved) partition function. For anti-chiral primaries, the
first two arguments of the partition function in the integrand
must be replaced by $q x^{{ 2}}$ and $\bar q\bar x^2$, respectively.
We know that this counting function for chiral primaries receives
contributions from the regular ones. The latter have been determined
above so that
\begin{equation} \label{ZNcp}
Z_{N}^{\cp}(q,\bar q) = Z_{N}^{\cp,\text{reg}}(q,\bar q) +
Z_{N}^{\cp,\text{exc}}(q,\bar q)
= T_N\left((q\bar q)^{1/6}\right) +
Z_{N}^{\cp,\text{exc}}(q,\bar q)\ .
\end{equation}
The counting function $T_N$ for regular chiral primaries has been
constructed in eqs. \eqref{TN} and \eqref{tTN} above. If all chiral
primaries were regular, there would be no additional contributions.
But we shall see below that this is not the case. Starting from $N=4$
not all chiral primaries are regular. The additional {\em exceptional}
chiral primaries are counted by $Z_N^{\cp,\text{exc}}$.

\subsection{Examples with $N \leq 3$: Minimal models}\label{secexample}

The aim of this and the following subsection is to illustrate our
general constructions through the first two examples, namely $N=2$
and $N=3$. These possess central charge $c_N < 3$ and hence they
belong to the minimal series of ${\cal N}=(2,2)$ superconformal
minimal models. For models from this series the chiral primaries
are well known. Our only task is therefore to show that the
general constructions of regular chiral primaries outlined in the
previous subsection allows us to recover all known chiral primaries.

\subsubsection*{$\mathbf{N=2}$}

Let us start by reviewing briefly the case of $N=2$ which gives
a CFT with Virasoro central charge $c_2 =1 \leq 3$. In section
3.3 we have listed all the sectors of this model along with the
conformal weight and maximal R-charge of their ground states.
From the results we can easily deduce that there are only two
sectors containing chiral primaries, namely the sectors
$({\text{\it id}}, [0])$ and  $({\text{\it v}},[2],1)$.
Recall that the label $({\text{\it v}},[2])$ labels a
branching function that can be decomposed into a sum of two
characters. These characters, which were displayed in eq.\
\eqref{brv2}, show that only one of the corresponding sectors
contains a chiral primary. Moreover, since the conformal
weight of all chiral primaries is bounded by $c_2/6 = 1/6$
there can be no chiral primaries among the excited states
of the model. Hence, we conclude that model contains two
chiral primaries. One is the identity field, the other one
a chiral primary of weight $h = 1/6$.

\begin{figure}
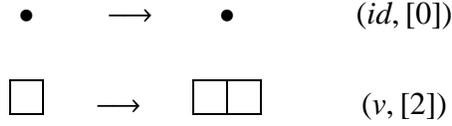

\begin{center}
\begin{tabular}{ccccc}
 $\bullet$ & \quad $\longrightarrow$ \quad  & $\bullet$ &\quad \quad  &
 $(\mbox{\it id},[0])$\\[6mm]
 \yng(1) & $\longrightarrow$ & \yng(2) & & $(\mbox{\it v},[2])$
\end{tabular}
\end{center}
\caption{The sets of Young diagrams $Y'$ and $Y$ for $N=2$.} \label{FigNeq2}
\end{figure}

Let us reproduce this simple conclusion from the construction
of regular chiral primaries. The construction we sketched above
instructs us to list all Young diagrams $Y'$ that can fit into
a rectangle of size $r' \times c'$ where $r' + c' = N = 2$.
Obviously, there are only two such Young diagrams, namely the
trivial one and the single box. These are depicted in the
leftmost column of figure~\ref{FigNeq2}. Applying the general
prescription \eqref{generation} (with $Y''\!=\!Y'$) we obtain
two Young diagrams $Y$ in the second column. From the two
columns we can read off the labels of the corresponding coset
sectors $(A,\{a\})$. These are displayed in the third column.
As we explained above, the first label $A$ is determined by
the number of boxes $n'$ of the Young diagram $Y'$ in the first
column. It is  $A =${\it id} if $n'$ is even and $A = v$
otherwise. The second entry contains the orbit $\{a\}$ of the
$\SU(2)_4$ representation $a$ that is associated with the Young
diagram $Y$ in the second column. According to eq.\ (\ref{cw}),
the conformal weights of the corresponding chiral primaries
are given by $h(\psi_{(A, a)}) = |Y'|/6$.
In this case, we recovered all chiral primaries through the
construction of the regular ones. The counting function for
chiral primaries is given by
\begin{align}
 Z_{2}^{\cp}(q,\bar q) = 1+ (q\bar q)^{1/6}\
 = T_2\left((q\bar q)^{1/6}\right)
\end{align}
and it obviously coincides with the counting function for
regular chiral primaries we stated in the previous subsection.

\subsubsection*{$\mathbf{N=3}$}

For $N=3$ we can proceed similarly. In this case, the model has
central charge $c_3 = 8/3$, still below the critical value $c=3$.
It is well known to possess 3 chiral primaries of conformal
weights $h = 0, 1/6, 1/3$ \cite{Gopakumar} and one can verify
this statement through a quick glance at the data we provided
with the list of necklaces in section 3.3. Note that the three
necklaces that are associated with chiral primaries have been
marked with the letters `CP' in the center.

\begin{figure}
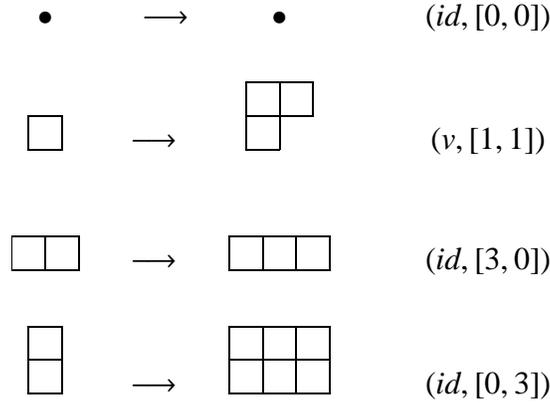

\begin{center}
\begin{tabular}{ccccc}
 $\bullet$ & \quad $\longrightarrow$ \quad  & $\bullet$ &\quad \quad  &
 $(\mbox{\it id},[0,0])$\\[6mm]
 \yng(1) & $\longrightarrow$ & \yng(2,1) & & $(\mbox{\it v},[1,1])$\\[10mm]
 \yng(2) & $\longrightarrow$ & \yng(3) & & $(\mbox{\it id},[3,0])$\\[6mm]
 \yng(1,1) & $\longrightarrow$ & \yng(3,3) & & $(\mbox{\it id},[0,3])$
\end{tabular}
\end{center}
\caption{The sets of Young diagrams $Y'$ and $Y$ for $N=3$.} \label{FigNeq3}
\end{figure}

Let us now apply our general construction of regular chiral primaries
to the case $N=3$. To begin with, we must list all the Young diagrams
$Y'$ that fit into a rectangle of size $1 \times 2$ or $2 \times 1$.
There are four such diagrams which are depicted in the leftmost
column of figure~\ref{FigNeq3}.  Application of the construction
\eqref{generation} (with $Y''\!=\!Y'$) gives four representations
of $\SU(3)$, as shown in the second column of figure~\ref{FigNeq3}.
The corresponding coset sectors are listed in the right column. In
this case two of the obtained sectors coincide since the $\SU(3)_6$
sectors $[0,3]$ and $[3,0]$ are related by the simple current
automorphism. Hence we end up with three inequivalent coset
representations whose ground states can provide a regular
chiral primary. Their labels are displayed in the third
column of figure~\ref{FigNeq3}.

We can easily scan the partition function $Z_3$ given in eq.\
(\ref{partfutN=3}) for chiral primaries from the list in the third
column of figure \ref{FigNeq3} to obtain
\begin{align}
Z_3^{\cp}(q,\bar q) = 1+ (q\bar q)^{1/6}+ (q\bar q)^{1/3}\ .
\end{align}
The answer agrees with the counting function for regular chiral
primaries we proposed in the previous subsection, i.e.\
$$ Z_3^{\cp}(q,\bar q) = Z_3^{\cp, \reg}(q,\bar q) = \tilde
T_3 ((q\bar q)^{1/6}) -(q\bar q)^{1/3}\ . $$
The subtraction of $(q\bar q)^{1/3}$ is explained by the field
identification $(id,[3,0])=(id,[0,3])$.

\subsection{Exceptional chiral primaries for $N \geq 4$}\label{secexceptional}

For $N=2,3$ the complete set of chiral primaries is given by the
class of regular chiral primaries. While this class still plays a
role for higher $N$ we shall find additional chiral primaries
when $N \geq 4$. We refer to  them as {\em exceptional} chiral
primaries.

\subsubsection*{$\mathbf{N=4}$}

For $N=4$ the central charge $c_4 = 5$ exceeds the bound $c=3$ that
can be reached with supersymmetric minimal models. Therefore we can no
longer rely on known results on the set of chiral primaries. Let us
therefore first apply our general constructions of regular chiral
primaries and then check whether they provide the complete set of
chiral primaries.

\begin{figure}[ht]
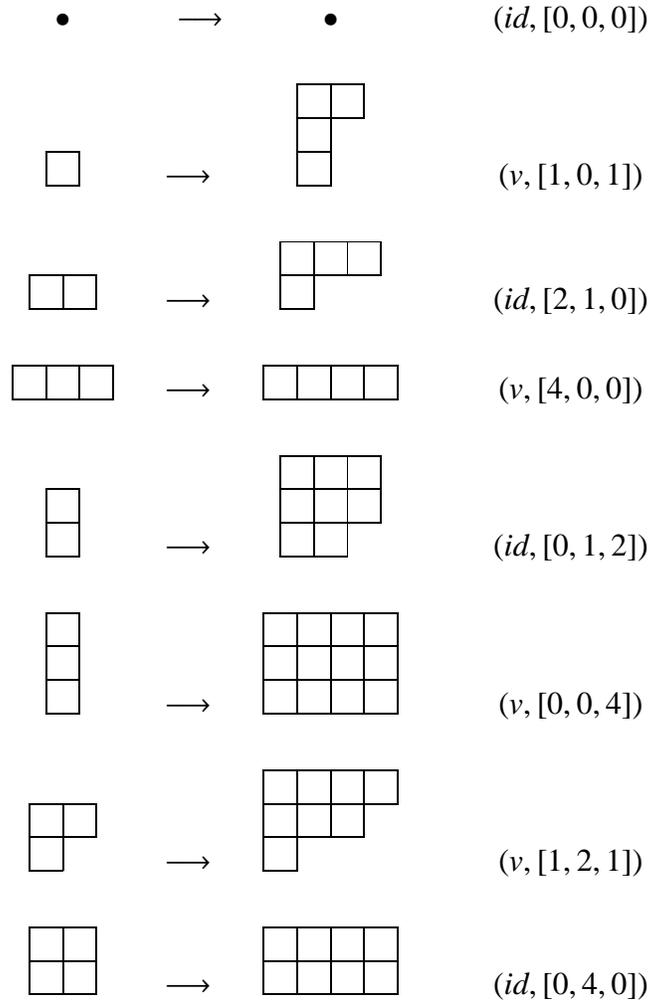

\begin{center}
\begin{tabular}{ccccc}
 $\bullet$ & \quad $\longrightarrow$ \quad  & $\bullet$ &\quad \quad  &
 $(\mbox{\it id},[0,0,0])$\\[6mm]
 \yng(1) & $\longrightarrow$ & \yng(2,1,1) & & $(\mbox{\it v},[1,0,1])$\\[6mm]
 \yng(2) & $\longrightarrow$ & \yng(3,1) & & $(\mbox{\it id},[2,1,0])$\\[6mm]
 \yng(3) & $\longrightarrow$ & \yng(4) & & $(\mbox{\it v},[4,0,0])$\\[6mm]
 \yng(1,1) & $\longrightarrow$ & \yng(3,3,2) & & $(\mbox{\it id},[0,1,2])$\\[6mm]
 \yng(1,1,1) & $\longrightarrow$ & \yng(4,4,4) & & $(\mbox{\it v},[0,0,4])$\\[6mm]
 \yng(2,1) & $\longrightarrow$ & \yng(4,3,1) & & $(\mbox{\it v},[1,2,1])$\\[6mm]
 \yng(2,2) & $\longrightarrow$ & \yng(4,4) & & $(\mbox{\it id},[0,4,0])$
 \end{tabular}
\end{center}
\caption{The sets of Young diagrams $Y'$ and $Y$ for $N=4$.} \label{constructionN=4}
\end{figure}

The analysis is summarized in figure~\ref{constructionN=4}. In the
first column we list all the Young diagrams $Y'$ which can fit
into rectangles of size $1 \times 3$ or $3 \times 1$ or $2 \times
2$. From these we build Young diagrams for representations of $\SU(4)$
with the help of eq.~\eqref{generation}. The results are shown in the
second column. Taking the first two columns together we determine the
list of coset sectors shown in the third column. Note that $(\mbox{\it
v},[4,0,0])$ and $(\mbox{\it v},[0,0,4])$ refer to the same sector of
the model since $[4,0,0]$ may be obtained from $[0,0,4]$ by applying
the simple current automorphism.  Hence, our construction gives
seven different coset sectors whose ground states are chiral primary.

In order to check whether we are missing any chiral primaries of the
model, we must scan the space of states with conformal weight $h \leq
c_4/6 = 5/6$, or a little less if we used the sector dependent bound
\eqref{bound}. Since $5/6 < 1$, all chiral primaries must be ground
states. Hence we can perform the scan by looking through the pairs
$(h,Q)$ we displayed when we listed the necklaces for $N=4$ in section
3.3. Those necklaces that give rise to chiral primaries have already
been marked by a `CP' in the center. Not surprisingly we find all the
seven regular chiral primaries from the third column of
figure~\ref{constructionN=4}.

On the other hand, the scan we just performed gives one more
chiral primary that does not appear in the right column of
figure~\ref{constructionN=4}, namely a ground state of the
coset sector $(id, [2,2,2])$. This new chiral primary has
conformal weight $h=Q=1/3$ and it is our first example of an
exceptional chiral primary. Our findings may be summarized in
the following expression
\begin{align}\label{resultZcp4}
 Z^{\cp}_4 (q, \bar q) &= 1 + (q\bar q)^{1/6} + 3 (q\bar q)^{1/3} + 2 (q \bar q)^{1/2}
+   (q\bar q)^{2/3}
\end{align}
which is equal to
\begin{align}
Z^{\cp}_4  (q, \bar q) &= Z^{\cp, \reg}_4  (q, \bar q) + (q\bar q)^{1/3} \,.\nn
\end{align}
The additional term $(q\bar q)^{1/3}$ counts the exceptional chiral primary.
A word of caution is in order. In general, chiral primaries can appear in
coset sectors which are fixed points of the theory. As we discussed before,
such fixed points must be resolved, and it is not {\em a priori} clear
whether this changes the multiplicity of the chiral primaries or not.  For
$N=4$ both the sector $(id, [0,4,0])$ and $(id,[2,2,2])$ are fixed
points and their ground states are chiral primary. The chiral
primaries appear with multiplicity one in both $\br^{\cal^W}_{(id+v,[0,4,0])}$
and $\br^{\cal W}_{(id+v,[2,2,2])}$, as can be seen from the their $q$-expansions
in appendix~\ref{appB2}. The result (\ref{resultZcp4}) for $Z_4^{\cp}$ holds true
provided that the fixed point resolution does {\em not} change the multiplicities. 
Otherwise the counting of chiral primaries would need to be modified accordingly.

\subsubsection*{$\mathbf{N=5}$}

As in the previous discussion we shall begin by listing all the
regular chiral primaries for $N=5$. After constructing all  Young
diagrams $Y'$ which fit into rectangles of size $2 \times 3$,
$3 \times 2$, $1 \times 4$ or $4 \times 1$ we apply
eq.~\eqref{generation} to obtain the 16 Young diagrams~$Y$. It
would take quite a bit of space to display all of them. So, let
us simply produce a list of the corresponding Dynkin labels,
\begin{align}
\S_5 = \{ &\repDid{[0,0,0,0]}, \repDv{[1,0,0,1]}, \repDid{[2,0,1,0]}, \repDid{[0,1,0,2]},\repDv{[0,0,1,3]}, \nn\\
&\repDv{[1,1,1,1]}, \repDv{[3,1,0,0]}, \repDid{[5,0,0,0]}, \repDid{[2,2,0,1]},  \repDid{[0,2,2,0]},\nn\\
& \repDid{[1,0,2,2]} , \repDid{[0,0,0,5]},  \repDv{[0,1,3,1]}, \repDv{[1,3,1,0]}, \repDid{[0,5,0,0]},\nn\\
& \repDid{[0,0,5,0]} \} \,. \nn
\end{align}
Note that both $\repDid{[5,0,0,0]}$ and $\repDid{[0,0,0,5]}$ as well as
$\repDid{[0,5,0,0]}$ and $\repDid{[0,0,5,0]}$ are identified by the simple
current automorphism. Hence we would expect 14 coset sectors whose
ground states are (regular) chiral primary.

Let us now look for the complete set of chiral primaries. In this case,
the sector independent bound \eqref{cbound} restricts the conformal
weight of chiral primaries to satisfy $h \leq c_5/6 = 4/3$. Here we
inserted the central charge $c_5 = 8$. One can again do a little better
using the sector dependent bound \eqref{bound}, but in the case at hand
we also listed all contributions to branching functions up to weight
$h = 4/3$, see appendix~\ref{appB3}. The results show that once more
all chiral primaries are ${\cal W}_N$ ground states so that we can
detect chiral primaries from the data that were provided in
section~\ref{secnecklaces} where we listed the necklaces for $N=5$.
We see 17 sectors of our ${\cal W}_N$ algebra contain a chiral primary
among its ground states. This is three more that the 14 regular
chiral primaries we described in the  previous paragraph. The
exceptional chiral primaries correspond to ground states of the
sectors $(v,[3,2,0,2])$, $(id,[2,3,1,1])$ and $(v, [2,2, 2, 2])$
and they possess conformal weights \mbox{$h=1/2$}, \mbox{$h=2/3$}
and \mbox{$h=5/6$}, respectively.

From the resolved partition function (\ref{partfuN=5}), we thus find
\begin{align}
 Z^{\cp}_5 &= 1 + (q\bar q)^{1/6} + 2 (q\bar q)^{1/3} + 4 (q \bar q)^{1/2}
+  5 (q\bar q)^{2/3} +  3 (q\bar q)^{5/6}+   q\bar q   \, \\[2mm]
    & = T_5\left( (q\bar q)^{1/6}\right) + (q\bar q)^{1/2} +
        (q\bar q)^{2/3} + (q\bar q)^{5/6}\ .
\end{align}
The last three terms give the counting function $Z_5^{\cp,\text{exc}}$
for exceptional chiral primaries.  Let us point out that one of the new
chiral primaries is sitting inside the fixed point sector $(v,a_\ast)$,
something that could not happen with the regular chiral primaries for
$N$ prime.

\setcounter{equation}{0}
\section{Conclusions}

In this paper we described the chiral symmetry ${\cal W}_N$ and
the complete modular invariant partition function $Z_N$ for a family
of field theories with ${\cal N}=(2,2)$ superconformal symmetry that
arise in the low energy limit of 1-dimensional adjoint QCD. We developed
techniques to study these theories for $N \geq 4$, where the theory does
not correspond to a supersymmetric minimal model. Special attention was
payed to the set of chiral primaries which are counted by a function
$Z_N^{\cp}(q)$ which we introduced in eq.\ \eqref{ZNcp}.

One of our main results is the discovery of exceptional chiral primaries
for $N=4,5$, which lie outside the set of regular chiral primaries. In
fact, we found one such chiral primary for $N=4$ and three of them for
$N=5$. Regular chiral primaries were described in some detail in section
\ref{secregular}. These fields are counted by a function $T_N(\q)$ which we
defined in eq.\ \eqref{TN}.

Our research is motivated by the desire to constrain the holographic
dual of the superconformal field theory under consideration. To this
end, one would like to find all chiral primaries in the limit of a
large number $N$ of colors. As we discussed in section 4, the
counting function $T_N$ for regular chiral primaries has a well-defined
and simple limiting behavior \eqref{TNlimit}. A similar analysis for the
exceptional chiral primaries has not been performed yet. It is possible
that such chiral primaries do not survive the large $N$ limit. We will
return to this problem in a forthcoming publication.

\section*{Acknowledgments}

We would like to thank Yuri Aisaka, Micha Berkooz, Terry Gannon,
Tigran Kalaydzhyan, \mbox{Andrey} Kormilitzin, Subir Sachdev and especially Kareljan
Schoutens for helpful discussions related to this work.  I.K.~thanks
Anton Nazarov for providing him an updated version of the mathematica
package affine.m \cite{Nazarov}.  The branching function computations
were performed on DESY's Theory Cluster and DESY's IT High Performance
Cluster (IT-HPC). This project was supported in parts by
the GIF grant no.\ 1038/2009.

\bigskip\bigskip\bigskip

\newpage

\section*{Appendix}

\appendix

\setcounter{equation}{0}
\section{Branching functions and fixed-point resolution for $N=2$} \label{appA}

In this appendix we explain how to compute the branching functions for
$N=2$, how to resolve the fixed point and how to recover the partition
function of a compactified free boson. The first subsection contains a
list of relevant functions and identities. Branching functions of the
model are computed in the second subsection before we discuss the
partition function in the final part.

\subsection{Notations}\label{appA1}

Throughout this appendix, we use the following notation for theta functions
\begin{align}
&\Psi_k(a|b)_x:=\sum_{n\in\mathbb{Z}} q^{\frac{a}{2}(n+\frac{1}{k})^2} x^{b(n+\frac{1}{k})} \nonumber \\
&\tilde{\Psi}_k(a|b)_x:=\sum_{n\in\mathbb{Z}} (-1)^n q^{\frac{a}{2}(n+\frac{1}{k})^2} x^{b(n+\frac{1}{k})} \,.\nonumber
\end{align}
For $k=1,2$, these reduce to ordinary Jacobi theta functions,
\begin{align}
&\theta_3(a|b)_x:=\Psi_1(a|b)_x\,,\quad \theta_4(a|b)_x:=\tilde{\Psi}_1(a|b)_x \,, \nn\\
& \theta_2(a|b)_x:=\Psi_2(a|b)_x\,,
\quad\theta_1(a|b)_x:=-i\tilde{\Psi}_2(a|b)_x. \nonumber
\end{align}
Whenever we set $x=1$, we omit the second parameter in the brackets
and the small subscript.

Next, let us introduce Ramanujan's theta function,
\begin{align}
f(a,b):= \sum_{n\in\mathbb{Z}}(ab)^{\frac{n^2}{2}}\left(\frac{a}{b}\right)^{\frac{n}{2}}=\prod_{n\in\mathbb{N}}\left(1+\frac{1}{a}(ab)^n\right)\left(1+\frac{1}{b}(ab)^n\right)\left(1-\frac{}{}(ab)^n\right). \nonumber
\end{align}
It is related to theta functions through
\begin{align}
  &f(a,b)=q^{-\frac{\beta^2}{8\alpha}}u^{-\frac{\beta}{2\alpha}}\Psi_{\frac{2\alpha}{\beta}}(\alpha|1)_u \nonumber\\
  &f(-a,-b)=q^{-\frac{\beta^2}{8\alpha}}u^{-\frac{\beta}{2\alpha}}\tilde{\Psi}_{\frac{2\alpha}{\beta}}(\alpha|1)_u\,
  \nonumber
\end{align}
where the variables $u$ and $q$ on the right hand side are related to $a$ and $b$
through $ab=q^{\alpha}$ and $a/b=q^{\beta}u^2$. Ramanujan's theta function obeys
Weierstrass' three-term relation \cite{Schwarz},
\begin{align}
f(a,b)f(c,d)=f(ad,bc)f(ac,bd)+a\,f(c/a,a^2 bd)f(d/a,a^2 bc)\,\nonumber
\end{align}
whenever $ab=cd$, and Hirschhorn's generalized quintuple product identity \cite{Hirschhorn},
\begin{align}
f(a,b)&f(c,d)=f(ac,bd)f(ac\cdot b^3,bd \cdot a^3)\nonumber\\&+a\,f\left(\frac{d}{a},\frac{a}{d}(abcd)\right)f\left(\frac{bc}{a},\frac{a}{bc}(abcd)^2\right)+
b\,f\left(\frac{c}{b},\frac{b}{c}(abcd)\right)f\left(\frac{ad}{b},\frac{b}{ad}(abcd)^2\right) \nonumber
\end{align}
for $(ab)^2=cd$. This concludes our brief list of mathematical functions
and identities.

\subsection{Branching functions}\label{appA2}
In order to illustrate how branching functions are computed, let us
focus on the decomposition of the ${\it id}+{\it v}$ sector of the
SO$(6)_1$ WZW model. According to our general prescription
\eqref{charSO1}, the corresponding character reads
\begin{align}
&\chi_{id+v}^{\text{SO}(6)_1} (q,x,z)\nn\\
&\hspace*{-3pt}=q^{-1/8} \prod_{n\in\mathbb{N}}(1+x^{1/3}q^{n-1/2})(1+x^{-1/3}q^{n-1/2})(1+x^{1/3}z^2q^{n-1/2}) \label{so(6)1}\nn\\
&\times(1+x^{-1/3}z^{-2}q^{n-1/2})(1+x^{1/3}z^{-2}q^{n-1/2})
(1+x^{-1/3}z^2q^{n-1/2})\\[2mm]
&\hspace*{-3pt}=\eta^{-3}f(q^{1/2}x^{1/3},q^{1/2}x^{-1/3})f(q^{1/2}x^{1/3}z^2,q^{1/2}x^{-1/3}z^{-2})
f(q^{1/2}x^{1/3}z^{-2},q^{1/2}x^{-1/3}z^2)\nonumber\\[2mm]
&\hspace*{-3pt}=\eta^{-3}f(q^{1/2}x^{1/3},q^{1/2}x^{-1/3})\left(f(q x^{2/3},q x^{-2/3})f(q z^{4},q z^{-4})+q^{1/2}x^{1/3}z^2f(q^2 x^{2/3}, x^{-2/3})f(q^2 z^{4}, z^{-4})\right)
\nn
\end{align}
where $\eta\equiv \eta(q)$ is the Dedekind eta function. In the final step  we
inserted Weierstrass' three-term relation. Using Hirschhorn's quintuple product,
one can then write
\begin{align}
&f(q^{1/2}x^{1/3},q^{1/2}x^{-1/3})f(qx^{2/3},qx^{-2/3})\nonumber\\
&~~~=f(q^{3/2}x,q^{3/2}x^{-1})f(q^3,q^3)+q^{1/2}f(q,q^5)\left(x^{1/3}f(q^{5/2}x,q^{1/2}x^{-1})+x^{-1/3}f(q^{5/2}x^{-1},q^{1/2}x)\right)\nn
\end{align}
for $a=q^{1/2}x^{1/3}$, $b=q^{1/2}x^{-1/3}$, $c=qx^{2/3}$, $d=qx^{-2/3}$ and
\begin{align}
&x^{1/3}f(q^{1/2}x^{1/3},q^{1/2}x^{-1/3})f(q^2x^{2/3},x^{-2/3})\nonumber\\
&~~~=q^{1/2}f(q^{3/2}x,q^{3/2}x^{-1})f(1,q^6)+f(q^2,q^4)\left(x^{1/3}f(q^{5/2}x,q^{1/2}x^{-1})+x^{-1/3}f(q^{5/2}x^{-1},q^{1/2}x)\right)\nn
\end{align}
for $a=q^{1/2}x^{1/3}$, $b=q^{1/2}x^{-1/3}$, $c=q^2x^{2/3}$, $d=x^{-2/3}$. In the second case we employed the following obvious symmetry property of $f$,
\begin{align*}
\frac{1}{\sqrt{u}}f\left(u,\frac{A}{u}\right)=\sqrt{u}f\left(Au,\frac{1}{u}\right).\nn
\end{align*}
Substituting these products back and simplifying, one arrives at
\begin{align}
&\chi_{id+v}^{\text{SO}(6)_1} (q,x,z)
\nn\\[2mm] &~~~=\eta^{-1}\Psi_1(3|1)_x \left(\frac{\theta_3(6)}{\eta^2}\theta_3(2|4)_z+ \frac{\theta_2(6)}{\eta^2} \theta_2(2|4)_z\right)\nonumber\\
&~~~~~~+ \eta^{-1}\left(\Psi_3(3|1)_x+\Psi_3(3|-1)_x\right) \left(\frac{\Psi_3(6)}{\eta^2}\theta_3(2|4)_z+\frac{\Psi_6(6)}{\eta^2} \theta_2(2|4)_z\right). \label{so(6)2}
\end{align}
In this formula $\theta_3(6)/\eta^{2}$, $\theta_2(6)/\eta^{2}$, $\Psi_3(6)/\eta^{2}$
and $\Psi_6(6)/\eta^{2}$ are combinations of SU$(2)_4$ string functions $c^0_0+c^0_4$, $2c^0_2$, $c^2_0$ and $c^2_2$, respectively (see e.g. \cite{KP} for more information). Expressions in round brackets can be then recognized as $S_{\{0\}}$ and $S_{\{2\}}$
which were defined in eq.\ \eqref{Sdef}. Indeed, taking into account the symmetry properties of the SU$(2)_k$ string functions,
$$c^{\Lambda}_{\lambda}=c^{J(\Lambda)}_{J(\lambda)}=
c^{\Lambda}_{\lambda+2k}=c^{\Lambda}_{-\lambda}\, \nonumber$$
where $J$  is the SU$(2)_k$ simple current, one readily sees
from
$$S_{\{a\}}^{N=2}=\sum_{\Lambda\in\{a\}}\sum_{\lambda\in\{-2,0,2,4\}}c^{\Lambda}_{\lambda}\sum_{n\in\mathbb{Z}}q^{4\left(n+\frac{\lambda}{8}\right)^2}z^{8\left(n+\frac{\lambda}{8}\right)}\nonumber$$ that
\begin{align}
&S_{\{0\}}=(c_0^0+c_4^0)\,\theta_3(2|4)_z+2c^0_2\, \theta_2(2|4)_z\nonumber\\
&S_{\{2\}}=c_0^2 \,\theta_3(2|4)_z+c^2_2\,\theta_2(2|4)_z.\nonumber
\end{align}
The decomposition (\ref{so(6)2}) of the SO$(6)_1$ character (\ref{so(6)1})
thus leads to the following branching functions
\begin{align}
&\br^{\cal W}_{(id+v,[0])}(q,x)=\eta^{-1} \Psi_1(3|1)_x \,,
\qquad \br^{\cal W}_{(id+v,[2])}(q,x)=\eta^{-1} \left(\Psi_3(3|1)_x+\Psi_3(3|-1)_x\right)
\,. \nonumber
\end{align}
Going along the same lines, one may compute the six remaining branching
functions. These are given by
\begin{equation}
\begin{aligned}\label{N=2chars}
&\br^{\cal W}_{(id-v,[0])}(q,x)=\eta^{-1}\tilde{\Psi}_1(3|1)_x \,,&\qquad& \br^{\cal W}_{(id-v,[2])}(q,x)=-\eta^{-1} \left(\tilde{\Psi}_3(3|1)_x+\tilde{\Psi}_3(3|-1)_x\right) \,,\\[2mm]
&\br^{\cal W}_{(sp+c,[0])}(q,x)=\eta^{-1}{\Psi}_2(3|1)_x  \,,&\qquad&\br^{\cal W}_{(sp+c,[2])}(q,x)=\eta^{-1}\left({\Psi}_6(3|1)_x+{\Psi}_6(3|-1)_x\right)\,, \\[2mm]
&\br^{\cal W}_{(sp-c,[0])}(q,x)=\eta^{-1}\tilde{\Psi}_2(3|1)_x \,,&\qquad& \br^{\cal W}_{(sp-c,[2])}(q,x)=-\eta^{-1}\left(\tilde{\Psi}_6(3|1)_x-\tilde{\Psi}_6(3|-1)_x\right)\,.
\end{aligned}
\end{equation}

\subsection{The partition function}\label{appA3}
The 'unresolved' partition function $\tilde Z_2$ of the \mbox{$N=2$}
model is constructed from the branching functions (\ref{N=2chars})
according to the general prescription (\ref{partfu}),
\begin{align}
\tilde Z_2 &=
   \sum_{A={\it id,v,sp,c}} \left(\vert \br^{\cal W}_{(A,[0])}\vert^2 +
\textstyle\frac{1}{2} \vert \br^{\cal W}_{(A,[2])}\vert^2 \right)=\frac{1}{2}\sum_{B={\it id+v,id-v,sp+c,sp-c}} \left(\vert \br^{\cal W}_{(B,[0])}\vert^2 +
\textstyle\frac{1}{2} \vert \br^{\cal W}_{(B,[2])}\vert^2 \right) \nonumber\\
&=\frac{1}{2|\eta|^2}\Big\{|\Psi_1(3|1)_x|^2+|\tilde\Psi_1(3|1)_x|^2+|\Psi_2(3|1)_x|^2+|\tilde\Psi_2(3|1)_x|^2+\frac{1}{2}[|\Psi_3(3|1)_x+\Psi_3(3|-1)_x|^2 \nonumber\\
&\quad\quad~+|\tilde\Psi_3(3|1)_x+\tilde\Psi_3(3|-1)_x|^2+|\Psi_6(3|1)_x+\Psi_6(3|-1)_x|^2+|\tilde\Psi_6(3|1)_x-\tilde\Psi_6(3|-1)_x|^2]\Big\} \,. \nn
\end{align}
As we explained before, the model suffers from a fixed point in the
sectors $(A,[2])$ so that $\tilde Z_2$ does not describe the partition
function of a well-defined CFT: The multiplicities of some states
inside the square brackets are non-integer. In order to cure the
issue, let us add the following modular-invariant contribution
\begin{align}
Z^{\rm res}_2&:=\frac{1}{4|\eta|^2} [|\Psi_3(3|1)_x-\Psi_3(3|-1)_x|^2
+|\tilde\Psi_3(3|1)_x-\tilde\Psi_3(3|-1)_x|^2\nonumber\\
&\qquad~~~+|\Psi_6(3|1)_x-\Psi_6(3|-1)_x|^2+|\tilde\Psi_6(3|1)_x+\tilde\Psi_6(3|-1)_x|^2]\,.
\end{align}
Note that this expression reduces to $Z^{\text res}_2(x=1) = 1$
due to Euler's pentagonal number theorem. Regrouping terms, we end up with
\begin{align}
Z_2&:=\tilde Z_2+ Z^{\rm res}_2\\
&=\frac{1}{2|\eta|^2}\Big\{|\Psi_1(3|1)_x|^2+|\tilde\Psi_1(3|1)_x|^2+
|\Psi_2(3|1)_x|^2+|\tilde\Psi_2(3|1)_x|^2+|\Psi_3(3|1)_x|^2+
|\Psi_3(3|-1)_x|^2\nonumber\\[2mm]
&\quad+|\tilde\Psi_3(3|1)_x|^2+|\tilde\Psi_3(3|-1)_x|^2+|\Psi_6(3|1)_x|^2+
|\Psi_6(3|-1)_x|^2+|\tilde\Psi_6(3|1)_x|^2+|\tilde\Psi_6(3|-1)_x|^2 \Big\} \,.\nn
\end{align}
With a little bit of additional effort, this expression may be resummed into a
more compact form
\begin{align}
&Z_2=\frac{1}{|\eta|^2}\sum_{n,w\in\mathbb{Z}}q^{\frac{k_L^2}{2}}\bar{q}^{\frac{k_R^2}{2}} x^{r k_L} \bar{x}^{r k_R}
  \qquad\text{with}\qquad k_{L,R}=\frac{n}{r}\pm\frac{w r}{2},\,\,\,\,\,\,\, r=2\!\sqrt{3}
\end{align}
which is the well known partition function of a free boson that has been
compactified on a circle of radius $r = 2\!\sqrt{3}$.

\setcounter{equation}{0}
\section{Branching functions  for $N=3,4,5$}\label{appB}

In this appendix we give the $q$-expansions of the branching functions
$\br^{\cal W}$ up to order $O(q^{c_N/6})$.  In order to better read off
the conformal weights $h$, we omit the overall factor $q^{-c_N/24}$ in
the $\br^{\cal W}$'s. As shown in
section~\ref{secbound}, there are no chiral primaries with conformal
weight larger than $h$=$c_N/6$. Chiral primary fields (with $h=Q$) are marked
by CP$_h$. We restrict to the NS sector, i.e.\ we only display the branching
functions $\br^{\cal W}_{(id+v,a)}$ ($a \in {\cal J}^0_N$). Similar expansions
exist for all the branching functions $\br^{\cal W}_{(sp+c,a)}$ in the R
sector.

\subsection{$\mathbf{N=3}$}\label{appB1}

The central charge is $c_3=8/3$. Chiral primaries exist only for $h
\leq 4/9$. The expansion of the branching functions $\br^{\cal
W}_{(id+v,\a)}$ ($a \in {\cal J}^0_3$) is given by
\begin{align}
\br^{\cal W}_{(id+ v,[0,0])}&= 1+O(q^1)
&&\text{\fbox{CP$_{0}$}  }\nn\\[2mm]
\br^{\cal W}_{(id+ v,[1,1])}&=  \left( x^{-1/3} + x^{1/3}\right) {q}^{1/6}+O(q^{2/3})
&&\text{\fbox{CP$_{1/6}$}  }\nn\\[2mm]
\br^{\cal W}_{(id+ v,[3,0])}&= \left( x^{-2/3} +1+ x^{2/3}\right) q^{1/3}+O(q^{5/6})
&&\text{\fbox{CP$_{1/3}$}  }\nn\\[2mm]
\br^{\cal W}_{(id+ v,[2,2])}&=  q^{1/9}+O(q^{11/18})\,.\nn
\end{align}

\subsection{$\mathbf{N=4}$}\label{appB2}

The central charge is $c_4=5$. The sector independent bound on the
conformal weight of a chiral primary state is therefore $h\leq5/6$.

The expansion of the branching functions $\br^{\cal W}_{(id+v,\a)}$
($a \in {\cal J}^0_4$) is given by
\begin{allowdisplaybreaks}
\begin{align}
&\br^{\cal W}_{(id+v,[0,0,0])}=1 + O({q}^{1})
&&\text{\fbox{CP$_{0}$}  }\nn\\
{}
&\br^{\cal W}_{(id+v,[1,0,1])}=   \left({x}^{1/3}+{x}^{-1/3}\right)q^{1/6}   + \left({x}^{2/3}+1+
 {x}^{-2/3}\right){q}^{2/3} + O({q}^{7/6})
&&\text{\fbox{CP$_{1/6}$}  }\nn\\[2mm]
{}
&\br^{\cal W}_{(id+v,[0,2,0])}={q}^{1/2}+O({q}^{1})\nn\\[2mm]
{}
&\br^{\cal W}_{(id+v,[2,1,0])}=\br^{\cal W}_{(id+v,[0,1,2])}= \left({x}^{2/3}+1+ {x}^{-2/3}\right) {{q}^{1/3}}
\nn\\[2mm]
&\qquad \qquad \qquad+ \left({x}+2 {x}^{1/3}+2{x}^{-1/3}+{x}^{-1}\right) {q}^{5/6}+O({q}^{4/3})
&&\text{\fbox{CP$_{1/3}$}  }\nn \\[2mm]
{}
&\br^{\cal W}_{(id+v,[4,0,0])}= \left({x}+{x}^{1/3}+{x}^{-1/3}+{x}^{-1}\right) {q}^{1/2}+O(q)
&&\text{\fbox{CP$_{1/2}$}  } \nn \\[2mm]
{}
&\br^{\cal W}_{(id+v,[2,0,2])}=q^{1/6} + \left({x}+2 {x}^{1/3}+2{x}^{-1/3}+{x}^{-1}\right){q}^{2/3}+O({q}^{7/6})\nn\\[2mm]
{}
&\br^{\cal W}_{(id+v,[1,2,1])}=\left({x}+2 {x}^{1/3}+2{x}^{-1/3}+{x}^{-1}\right){q}^{1/2} +O({q})
&&\text{\fbox{CP$_{1/2}$}  }\nn\\[2mm]
{}
&\br^{\cal W}_{(id+v,[2,3,0])}=
 \left({x}^{2/3}+1+{x}^{-2/3}\right){q}^{1/2}+O({q})
\nn\\[2mm]
{}
&\br^{\cal W}_{(id+v,[3,1,1])}= \left({x}^{1/3}+{x}^{-1/3}\right) {{q}^{1/4}}+O({q}^{5/4})
\nn\\[2mm]
{}
&\br^{\cal W}_{(id+v,[0,4,0])}=  \left(x^{4/3}+x^{2/3}+2 + x^{-2/3}+x^{-4/3} \right)q^{2/3}+O({q}^{7/6})
&&\text{\fbox{CP$_{2/3}$}  }\nn\\[2mm]
{}
&\br^{\cal W}_{(id+v,[2,2,2])}=  \left( x^{2/3}+2 + x^{-2/3} \right){q}^{1/3}\nn\\
&\qquad\qquad\qquad+\left( 3 x+5 x^{1/3} + 5x^{-1/3} + 3 x^{-1} \right) {q}^{5/6}+O({q}^{4/3})
&&\text{\fbox{CP$_{1/3}$}  }\nn \,.
\end{align}
\end{allowdisplaybreaks}

\subsection{$\mathbf{N=5}$}\label{appB3}

The central charge is $c_5=8$, and we expand up to order $O(q^{4/3})$ in order
to capture all contributions from chiral primaries with conformal weight
$h \leq c_5/6$. In order to write the expansion of branching functions
$\br^{\cal W}_{(id+v,\a)}$ ($a \in {\cal J}^0_5$) we shall introduce the
shorthand $y^n \equiv x^{n} + x^{-n}$.
\begin{allowdisplaybreaks}
\begin{align}
\br^{\cal W}_{(id+v,[0,0,0,0])}&=1+q +O(q^{3/2})&&\text{\fbox{CP$_{0}$}  }\nn\\[2mm]
\br^{\cal W}_{(id+v,[1,0,0,1])}&=y^{ 1/3} q^{1/6}
+(y^{ 2/3}+1 ) \, q^{2/3}
+(y+3 y^{ 1/3} )\, q^{7/6}+O(q^{5/3})&&
\text{\fbox{CP$_{1/6}$}  }\nn\\[2mm]
\br^{\cal W}_{(id+v,[0,1,1,0])}&=q^{7/15}+(y+2 y^{ 1/3} ) \, q^{29/30}+O(q^{22/15})
\nn\\[2mm]
\br^{\cal W}_{(id+v,[2,0,1,0])}&=(y^{ 2/3}+1) \, q^{1/3}
+ (y+2 y^{ 1/3}) \,q^{5/6}  \nn\\[2mm]
&~~~+(y^{ 4/3}+ 5 y^{ 2/3}+6) \,q^{4/3}\ +O(q^{11/6})&&
\text{\fbox{CP$_{1/3}$}}\nn\\[2mm]
\br^{\cal W}_{(id+v,[1,2,0,0])}&= y^{ 1/3}  q^{7/10}  +(y^{ 4/3}+ 3 y^{ 2/3}+4 )\,  q^{6/5}
+O(q^{17/10})&&
\nn\\[2mm]
\br^{\cal W}_{(id+v,[0,1,0,2])}&=\br^{\cal W}_{(id+v,[2,0,1,0])}&&
\text{\fbox{CP$_{1/3}$}}\nn\\[2mm]
\br^{\cal W}_{(id+v,[0,0,2,1])}&=\br^{\cal W}_{(id+v,[1,2,0,0])}&&
\nn\\[2mm]
\br^{\cal W}_{(id+v,[3,1,0,0])}&=(y+y^{ 1/3})\, q^{1/2} +( y^{ 4/3}+2 y^{ 2/3}+3)\,  q
+O(q^{3/2})&&
\text{\fbox{CP$_{1/2}$}  }\nn\\[2mm]
\br^{\cal W}_{(id+v,[2,0,0,2])}&={q^{1/5}}+(y+2 y^{ 1/3})\, q^{7/10}+ (2 y^{ 4/3}+4 y^{ 2/3}+7)\, q^{6/5}
\!+\!O(q^{17/10})\!\!\!\!\!
\nn\\[2mm]
\br^{\cal W}_{(id+v,[1,1,1,1])}&= (y+2 y^{ 1/3} ) \,q^{1/2} +(2 y^{ 4/3}+6 y^{ 2/3}+8)\, q
+O(q^{3/2})&&
\text{\fbox{CP$_{1/2}$}  }\nn\\[2mm]
\br^{\cal W}_{(id+v,[1,0,3,0])}&=\br^{\cal W}_{(id+v,[0,3,0,1])}
&&
\nn\\[2mm]
\br^{\cal W}_{(id+v,[0,3,0,1])}&= (y^{ 2/3}+1)\, q^{4/5}  + ( y^{ 5/3}+4 y+6y^{ 1/3}) \, q^{13/10}
+O(q^{9/5})&&
\nn\\[2mm]
\br^{\cal W}_{(id+v,[0,2,2,0])}&=(y^{ 4/3}+y^{ 2/3}+2) \, q^{2/3}+ ( y^{ 5/3}+3 y+5 y^{ 1/3}) \, q^{7/6}
+O(q^{5/3})&&
\text{\fbox{CP$_{2/3}$}  }\nn\\[2mm]
\br^{\cal W}_{(id+v,[0,0,1,3])}&=\br^{\cal W}_{(id+v,[3,1,0,0])}&&
\text{\fbox{CP$_{1/2}$}  }\nn\\[2mm]
\br^{\cal W}_{(id+v,[5,0,0,0])}&= (y^{ 4/3}+y^{ 2/3}+1)\, q^{2/3}  +  ( y^{ 5/3}+ y+2y^{ 1/3})\, q^{7/6}
+O(q^{5/3})&&
\text{\fbox{CP$_{2/3}$}  }\nn\\[2mm]
\br^{\cal W}_{(id+v,[3,0,1,1])}&=  y^{ 1/3} {q}^{3/10} + (y^{ 4/3}+3 y^{ 2/3}+4)\, q^{4/5}  \nn\\[2mm]
&~~~+ (2 y^{ 5/3}+7 y+13y^{ 1/3})\, q^{13/10}
+O(q^{41/30})\!\!\!\! &&
\nn\\[2mm]
\br^{\cal W}_{(id+v,[2,2,0,1])}&= (y^{ 4/3}+ 2 y^{ 2/3}+3 )\, q^{2/3}  +( 2 y^{ 5/3}+6 y+10 y^{ 1/3})\,  q^{7/6}   +O(q^{5/3})
\!\!\!&&
\text{\fbox{CP$_{2/3}$}  }\nn\\[2mm]
\br^{\cal W}_{(id+v,[2,1,2,0])}&=(y^{ 2/3}+1)\, q^{8/15} +( y^{ 5/3}+4 y+7y^{ 1/3})\, q^{31/30}
+O(q^{23/15})&&
\nn\\[2mm]
\br^{\cal W}_{(id+v,[1,3,1,0])}&=(  y^{ 5/3}+ 2 y +3 {y}^{1/3} )\, q^{5/6} \nn\\[2mm]
&~~~+ (2 y^{ 2}+6 y^{ 4/3}+11 y^{ 2/3}+13)\, {q}^{4/3}
+O(q^{11/6})&&
\text{\fbox{CP$_{5/6}$}  }\nn\\[2mm]
\br^{\cal W}_{(id+v,[1,1,0,3])}&=\br^{\cal W}_{(id+v,[3,0,1,1])}
&&
\nn\\[2mm]
\br^{\cal W}_{(id+v,[1,0,2,2])}&=\br^{\cal W}_{(id+v,[2,2,0,1])}
&&
\text{\fbox{CP$_{2/3}$}  }\nn\\[2mm]
\br^{\cal W}_{(id+v,[0,5,0,0])}&= (y^{ 2}+y^{ 4/3} +2 y^{ 2/3}+2)\, q  +O(q^{3/2})
&&\text{\fbox{CP$_{1}$}  }\nn\\[2mm]
\br^{\cal W}_{(id+v,[0,2,1,2])}&= (y^{ 2/3}+1)\, {q}^{8/15}  + ( y^{ 5/3}+4 y+7y^{ 1/3} )\, {q}^{31/30}
+ O(q^{23/15})&&
\nn\\[2mm]
\br^{\cal W}_{(id+v,[0,1,3,1])}&=\br^{\cal W}_{(id+v,[1,3,1,0])}
&&
\text{\fbox{CP$_{5/6}$}  }\nn\\[2mm]
\br^{\cal W}_{(id+v,[4,1,0,1])}&= (y^{ 2/3}+1)\, {q}^{2/5} + (y^{ 5/3}+3y+5 {y}^{ 1/3})\, {q}^{9/10}
+O(q^{7/5})&&
\nn\\[2mm]
\br^{\cal W}_{(id+v,[4,0,2,0])}&={q}^{4/15} +  (y+2 y^{ 1/3})\,  {q}^{23/30}   +  (3 y^{ 4/3}+6 y^{ 2/3}+10)\,   {q}^{19/15}   +O(q^{7/5})
\!\!\!\!\!\!\!
\nn\\[2mm]
  \br^{\cal W}_{(id+v,[3,2,1,0])}&= (y+ 2 y^{ 1/3} )  \, q^{19/30}+  (y^{ 2}+4 y^{ 4/3}+9 y^{ 2/3}+11)\, q^{17/15}
+O(q^{49/30})
\nn\\[2mm]
\br^{\cal W}_{(id+v,[3,0,0,3])}&=
(y^{ 2/3}+1) \, q^{3/5} +   (y^{ 5/3}+3 y+5 {y}^{ 1/3})\, q^{11/10}
+O(q^{8/5})
\nn\\[2mm]
\br^{\cal W}_{(id+v,[2,4,0,0])}&= (y^{ 4/3}+y^{ 2/3}+2 )\, q^{13/15}+O(q^{41/30})
\nn\\[2mm]
\br^{\cal W}_{(id+v,[2,1,1,2])}&= (  y^{ 2/3}+2) \, q^{2/5} + (y^{ 5/3}+5 y+9 {y}^{ 1/3})\,q^{9/10}+O(q^{7/5})
\nn\\[2mm]
\br^{\cal W}_{(id+v,[2,0,3,1])}&= ( y+ 2 y^{ 1/3})\,  q^{7/10}+ (y^{ 2}+5y^{ 4/3}+10y^{ 2/3}+13)\, q^{6/5}+O(q^{17/10})
\nn\\[2mm]
\br^{\cal W}_{(id+v,[1,3,0,2])}&=\br^{\cal W}_{(id+v,[2,0,3,1])}
\nn\\[2mm]
\br^{\cal W}_{(id+v,[1,2,2,1])}&= ( y+ 2 y^{ 1/3}) \, q^{17/30}+ (y^{ 2}+5 y^{ 4/3}+11 y^{ 2/3}+14)\, q^{16/15}+O(q^{47/30})
\!\!\!\!\!\!\!&&
\nn\\[2mm]
\br^{\cal W}_{(id+v,[1,1,4,0])}&=( y^{ 4/3}+2 y^{ 2/3}+2 ) \, q^{4/5}\nn\\[2mm]
&~~~+(y^{ 7/3}+4 y^{ 5/3}+9 y+13 {y}^{ 1/3})\, q^{13/10}+O(q^{9/5})
\nn\\[2mm]
\br^{\cal W}_{(id+v,[0,3,3,0])}&= q^{3/5}+ (y^{ 5/3}+3 y+5 {y}^{ 1/3})\, q^{11/10}+O(q^{8/5})
\nn\\[2mm]
\br^{\cal W}_{(id+v,[0,1,2,3])}&=\br^{\cal W}_{(id+v,[3,2,1,0])}
\nn\\[2mm]
\br^{\cal W}_{(id+v,[3,2,0,2])}&= ( y+ 2 y^{ 1/3})\,  q^{1/2}+ (3 y^{ 4/3}+7 y^{ 2/3}+9 )\, q+O(q^{3/2})
&&
\text{\fbox{CP$_{1/2}$}  }\nn\\[2mm]
\br^{\cal W}_{(id+v,[3,1,2,1])}&=y^{ 1/3} q^{11/30}+ (2 y^{ 4/3}+6 y^{ 2/3}+8 )\, q^{13/15}+O(q^{41/30})
\nn\\[2mm]
\br^{\cal W}_{(id+v,[2,3,1,1])}&=(y^{ 4/3} + 3 y^{ 2/3}+ 4)\, q^{2/3}\nn\\[2mm]
&~~~+ (4 y^{ 5/3}+12 y+19  {y}^{ 1/3}) \, q^{7/6}+O(q^{5/3})
&&
\text{\fbox{CP$_{2/3}$}  }\nn\\[2mm]
\br^{\cal W}_{(id+v,[2,2,3,0])}&=(y^{ 2/3}+1)\, q^{7/15}+ (y^{ 5/3}+4 y+7 {y}^{ 1/3} )\, q^{29/30}+O(q^{22/15})
\nn\\[2mm]
\br^{\cal W}_{(id+v,[2,2,2,2])}&= q^{1/3}+( y^{ 5/3}+5 y +9 {y}^{ 1/3})  \, q^{5/6}\nn\\[2mm]
&~~~+ (4 y^{ 2}+19 y^{ 4/3}+36 y^{ 2/3}+47  )\, q^{4/3} + O(q^{11/16})\nn
&&\text{\fbox{CP$_{5/6}$}  } \,.
\end{align}
\end{allowdisplaybreaks}

\setcounter{equation}{0}
\section{$\SU(N)_{2N}$ representations with zero monodromy charge}\label{appC}

In this appendix we will prove the formula (\ref{C2generalized}) for
the quadratic Casimir $C_2(Y)$ of a representation $a \in {\cal
J}^0_N$ associated with a Young diagram $Y$. As described in section
3 we pick up a pair of SU$(N)$ Young diagrams $Y'$ and $Y''$ satisfying
the conditions listed in the first paragraph of section~\ref{integersec}.
From these two Young diagrams we build a new diagram $Y = (l_1, \dots, l_{N-1})$
through our prescription \eqref{generation}.

We now claim that the resulting Young diagram $Y$ possesses
\begin{align}
|Y|=r'' N \label{num-boxes}
\end{align}
boxes and that the eigenvalue of the SU$(N)$ quadratic Casimir
on $Y$ takes the value
\begin{align}
C_2(Y) = n' N + C_2(Y') - C_2(Y'') \,. \label{C2generalized2}
\end{align}
In order to prove these two statements, we use eq.\ (\ref{generation})
to obtain
\begin{align}
|Y|&=\sum_{i} l_i=\sum_{i=1}^{r'}(r''+l_i')+r''(N-r'-l_1'')+\sum_{i=1}^{r''-1}(r''-i)(l_i''-l_{i+1}'') \,.
\end{align}
Since
\begin{align}
\sum_{i=1}^{r''-1}i(l_i''-l_{i+1}'')=\sum_{i=1}^{r''-1}i\,l_i''-\sum_{i=2}^{r''}(i-1)l_i''=n'-r''l''_{r''} \label{identity-yt}\,,
\end{align} we arrive at
\begin{align}
|Y|&=r'r''+n'+r''(N-r'-l_1'')+r''(l_1''-l_{r''}'')-(n'-r''l_{r''}'')=r''N \,,
\end{align}
which proves eq.\ (\ref{num-boxes}).

The quadratic Casimir on $Y$ is therefore given by
\begin{align}
C_{2}(Y)=\frac{1}{2} \left[ N r''(N+1-r'') + \sum_{i}  l_i(l_i-2 i) \right]. \nonumber
\end{align}
Let us compute the last term in the brackets,
\begin{allowdisplaybreaks}
\begin{align}
\sum_{i} & l_i(l_i-2 i) \nonumber\\
&=\sum_{i=1}^{r'} (r''+l_i')(r''+l_i'-2i)+\sum_{i=1}^{N-l_1''-r'} r''(r''-2(r'+i)) \nonumber\\
&~~~+\sum_{i=1}^{l_1''-l_2''} (r''-1)(r''-1-2(N-l_1''+i)+\cdots+\sum_{i=1}^{l_{r''-1}''-l_{r''}''} 1\cdot (1-2(N-l_{r''-1}''+i)) \nonumber\\
&=\sum_{i=1}^{r'} (r''+l_i')(r''+l_i'-2i)-\sum_{i=1}^{N-l_1''-r'} r''(r''+2i)+\sum_{i=1}^{r''-1}(r''-i)(r''-i-2N)(l_i''-l_{i+1}'') \nonumber\\
&~~~+2\sum_{i=1}^{r''-1}(r''-i)l_i''(l_i''-l_{i+1}'')-\sum_{i=1}^{r''-1}(r''-i)(l_i''-l_{i+1}''+1)(l_i''-l_{i+1}'') \nonumber\\
&=:\Sigma_1+\Sigma_2+\Sigma_3+\Sigma_4+\Sigma_5 \,.
\end{align}
\end{allowdisplaybreaks}
Using the identities analogous to eq.\ (\ref{identity-yt}),
\begin{equation}
\begin{aligned}
&\sum_{i=1}^{r''-1}i(l_i''^2-l_{i+1}''^2)
=\sum_{i=1}^{r''} l_i''^2-r''l_{r''}''^2 \,, \\
&\sum_{i=1}^{r''-1}i^2(l_i''-l_{i+1}'')
=2\sum_{i=1}^{r''}il_i''-n'-(r''^2-r'')l_{r''}'' \,,
\end{aligned}
\end{equation}
we can conclude
\begin{equation}
\begin{aligned}
\Sigma_1+\Sigma_2&=\sum_{i=1}^{r'}l_i'(l_i'-2i)+2r''n'-r''(N-l_1'')(N-l_1''-r''+1) \,,\\
\Sigma_3&=2\sum_{i=1}^{r''}il_i''+(2N-2r''-1)n'+(r''^2-2r''N)l_1'' \,,\\
\Sigma_4+\Sigma_5&=\sum_{i=1}^{r''-1}(r''-i)((l_i''^2-l_{i+1}''^2)-(l_i''-l_{i+1}''))\\
&=-\sum_{i=1}^{r''}l_i''^2+n'+r''(l_1''^2-l_1'') \,.
\end{aligned}
\end{equation}
When summed up, this contributions give
\begin{align}
\sum_{i} l_i(l_i-2 i)&=-Nr''(N+1-r'')+2n' N+\sum_{i=1}^{r'} l_i'(l_i'-2 i)-\sum_{i=1}^{r''} l_i''(l_i''-2 i)
\end{align}
and thus
\begin{align}
C_{2}(Y)&=Nn'+\frac{1}{2}\left(\sum_{i=1}^{r'} l_i'(l_i'-2 i)-\sum_{i=1}^{r''} l_i''(l_i''-2 i)\right) \,.
\end{align}
Since $|Y'|= |Y''|=n'$ holds by construction, this expression is equivalent
to eq.\ (\ref{C2generalized2}).


\end{document}